\documentstyle[12pt]{article}

\begin{document}

\baselineskip=14pt plus 1pt minus 1pt

\begin{center}{\large \bf
$\Delta I=2$ staggering 
in rotational bands of diatomic molecules as a manifestation of 
interband interactions}

\bigskip\bigskip

{Dennis Bonatsos$^{\#}$,
C.~Daskaloyannis$^+$,
S. B. Drenska$^\dagger$,
N. Karoussos$^{\#}$,
J. Maruani$^*$, 
N. Minkov$^\dagger$, 
P. P. Raychev$^\dagger$,
R. P. Roussev$^\dagger$
\bigskip

{$^{\#}$ Institute of Nuclear Physics, N.C.S.R.
``Demokritos''}

{GR-15310 Aghia Paraskevi, Attiki, Greece}

{$^+$ Department of Physics, Aristotle University of
Thessaloniki}

{GR-54006 Thessaloniki, Greece }

{$^\dagger$ Institute for Nuclear Research and Nuclear Energy, Bulgarian
Academy of Sciences }

{72 Tzarigrad Road, BG-1784 Sofia, Bulgaria}}

{$^*$ Laboratoire de Chimie Physique, CNRS and UPMC}

{11, rue Pierre et Marie Curie, F-75005 Paris, France} 

\end{center}

\bigskip\bigskip
\centerline{\bf Abstract} \medskip
It is  shown that the recently observed
$\Delta I=2$ staggering seen in superdeformed nuclear bands
is also occurring in certain electronically excited 
rotational bands of diatomic molecules.
In the case of diatomic molecules the effect is attributed to interband
interactions (bandcrossings).

\bigskip\bigskip
PACS numbers: 33.20.Sn, 21.10.Re, 21.60.Ev

\newpage

{\bf I. Introduction} 

Several {\sl staggering} effects are known in nuclear spectroscopy
\cite{BM}: 

1) In rotational $\gamma$ bands of even nuclei the energy levels with 
odd angular momentum $I$ ($I$=3, 5, 7, 9, \dots) are slightly displaced 
relatively to the levels with even $I$ ($I$=2, 4, 6, 8, \dots), i.e. 
the odd levels do not lie at the energies predicted by an $E(I)=A I(I+1)$ fit 
to the even levels, but all of them lie systematically above or all of 
them lie systematically below the predicted energies \cite{PLB200}. 

2) In octupole bands of even nuclei the levels with odd $I$ and negative 
parity ($I^{\pi}$=1$^-$, 3$^-$, 5$^-$, 7$^-$, \dots) are displaced relatively 
to the levels with even $I$ and positive parity ($I^{\pi}$=0$^+$, 2$^+$, 
4$^+$, 6$^+$, \dots) \cite{Phill}. 

3) In odd nuclei, rotational bands (with $K=1/2$) separate into 
signature partners, i.e. the levels with $I$=3/2, 7/2, 11/2, 15/2, \dots
are displaced relatively to the levels with $I$=1/2, 5/2, 9/2, 13/2, \dots
\cite{WuZhou}.   

In all of the above mentioned cases each level with angular momentum $I$ 
is displaced relatively to its neighbours with angular momentum $I\pm 1$.
The effect is then called {\sl $\Delta I=1$ staggering}.

A new kind of staggering ({\sl $\Delta I=2$ staggering}) has been recently 
observed \cite{Fli,Ced} in superdeformed nuclear bands 
\cite{Twin,Nolan,Janssens}. 
If $\Delta I=2$ staggering is present,
then, for example, the levels with $I$=2, 6, 10, 14, \dots are displaced 
relatively to the levels with $I$=0, 4, 8, 12, \dots, i.e. the level with 
angular momentum $I$ is displaced relatively to its neighbours with 
angular momentum $I\pm 2$. 

Although $\Delta I=1$ staggering of the types mentioned above has been 
observed in several nuclei and certainly is an effect larger than the 
relevant experimental uncertainties, $\Delta I=2$ staggering has been seen 
in only a few cases \cite{Fli,Ced,Semple,Kruecken}
and, in addition, the effect is not clearly larger 
than the relevant experimental errors. 

There have been by now several theoretical works related to the 
possible physical origin of the $\Delta I=2$ staggering effect
\cite{SZG,MQ,Mag,Kota,Liu,Pav,Wu}, some of 
them \cite{HM,Macc,PavFli,Doenau,Luo,Magi}
using symmetry arguments which could be of applicability to other 
physical systems as well. 

On the other hand, rotational spectra of diatomic molecules 
\cite{Herz} are known to 
show great similarities to nuclear rotational spectra, having in addition 
the advantage that observed  rotational bands in several diatomic molecules 
\cite{YD,CrD,CrH,CoH}
are much longer than the usual rotational nuclear bands. We have been 
therefore motivated to make a search for $\Delta I =2$ staggering in 
rotational bands of diatomic molecules. The questions to which we have hoped
to provide answers are:

1) Is there $\Delta I =2$ staggering in rotational bands of diatomic 
molecules? 

2) If there is, what are its possible physical origins?  

In Section II of the present work the $\Delta I=2$ staggering in
superdeformed 
nuclear bands will be briefly reviewed. Evidence from existing 
experimental data for $\Delta I=2$ staggering in rotational bands 
of diatomic molecules will be presented in Section III and discussed in 
Section IV, while Section V will contain the conclusions drawn. 

\bigskip
{\bf II. $\Delta I=2$ staggering in superdeformed nuclear bands}
\medskip

In nuclear physics the experimentally determined quantities are the
$\gamma$-ray transition energies between levels differing by two units
of angular momentum ($\Delta I=2$). For these the symbol
\begin{equation}  
E_{2,\gamma}(I) = E(I+2)-E(I) 
\end{equation} 
is used, where $E(I)$ denotes the energy of the level with angular momentum
$I$.
The deviation of the $\gamma$-ray transition energies from the
rigid rotator behavior can be measured by the quantity \cite{Ced}
\begin{equation}
 \Delta E_{2,\gamma}(I) = {1\over 16} (6E_{2,\gamma}(I) -4E_{2,\gamma} (I-2)
-4E_{2,\gamma}(I+2) +E_{2,\gamma}(I-4) +E_{2,\gamma}(I+4)). 
\end{equation} 

\noindent Using the rigid rotator expression 
\begin{equation}
E(I)=A I(I+1),
\end{equation} 
one can easily see that
in this case $\Delta E_{2,\gamma} (I) $ vanishes.
In addition the perturbed rigid rotator expression 
\begin{equation}
E(I)= A I(I+1) + B (I(I+1))^2,
\end{equation}
 gives vanishing $\Delta E_{2,\gamma} (I)$. 
These properties are due to the fact that Eq. (2) is a (normalized) 
discrete approximation of the fourth derivative of the function
$E_{2,\gamma}(I)$, i.e. essentially the fifth derivative of the 
function $E(I)$. 

In superdeformed nuclear bands the angular momentum of the observed states 
is in most cases unknown. To avoid this difficulty, the quantity 
$\Delta E_{2,\gamma}$ is usually plotted not versus the angular momentum $I$, 
but versus the angular frequency 
\begin{equation} 
\omega = {dE(I)\over dI}, 
\end{equation}
which for discrete states takes the approximate form 
\begin{equation}
\omega = {E(I+2)-E(I)\over \sqrt{(I+2)(I+3)}-\sqrt{I(I+1)} }.
\end{equation}
For large $I$ one can take the Taylor expansions of the square roots in 
the denominator, thus obtaining
\begin{equation}
\omega = {E(I+2)-E(I) \over 2} = {E_{2,\gamma}(I) \over 2}.
\end{equation} 

Examples of superdeformed nuclear bands exhibiting staggering are shown in 
Figs 1--2 \cite{Fli,Ced}. We say that $\Delta I=2$ staggerimg is observed if 
the quantity $\Delta E_2(I)$ exhibits alternating signs with increasing 
$\omega$ (i.e. with increasing $I$, according to Eq. (7)). The following 
observations can be made:

1) The magnitude of $\Delta E_2(I)$ is of the order of 10$^{-4}$--10$^{-5}$
times the size of the gamma transition energies. 

2) The best example of $\Delta I=2$ staggering is given by the first 
superdeformed band of $^{149}$Gd, shown in Fig. 1a. In this case the effect 
is almost larger than the experimental error.

3) In most cases the $\Delta I=2$ staggering is smaller than the experimental 
error (see Figs 1b, 2a, 2b), with the exception of a few points in Fig. 1b. 

\bigskip
{\bf III. $\Delta I=2$ staggering in rotational bands of diatomic molecules}
\medskip

In the case of molecules \cite{Bar} the experimentally determined
quantities regard the R branch ($(v_{lower},I)\rightarrow (v_{upper},I+1)$) 
and the P branch ($(v_{lower},I)\rightarrow (v_{upper},I-1)$), where
$v_{lower}$ is the vibrational quantum number of the initial state, 
while $v_{upper}$ is the vibrational quantum number of the final state. 
They are related to transition energies through the equations \cite{Bar}
\begin{equation}
E^R(I)-E^P(I)= E_{v_{upper}} (I+1) -E_{v_{upper}} (I-1) = 
DE_{2, v_{upper}} (I), 
\end{equation} 
\begin{equation}
E^R(I-1)-E^P(I+1) = E_{v_{lower}}(I+1)-E_{v_{lower}}(I-1)=
DE_{2, v_{lower}}(I),
\end{equation} 
where in general
\begin{equation}
 DE_{2,v} (I) = E_v(I+1)-E_v(I-1).
\end{equation} 
$\Delta I=2$ staggering  can then
be estimated by using Eq. (2), with $E_{2,\gamma}(I)$ replaced by 
$DE_{2,v}(I)$:
\begin{equation}
 \Delta E_{2,v} (I)= {1\over 16} (6 DE_{2,v}(I)-4 DE_{2,v}(I-2)
-4 DE_{2,v}(I+2) +DE_{2,v}(I-4) +DE_{2,v}(I+4)). 
\end{equation} 

Results for several rotational bands in different electronic and vibrational 
states of various diatomic molecules are shown in Figs 3--9. 
We say that $\Delta I=2$ staggering is observed if the quantity 
$\Delta E_2(I)$ exhibits alternating signs with increasing $I$ ($I$ is 
increased by 2 units each time). The magnitude of $\Delta E_2(I)$ is
usually of the order of 10$^{-3}$--10$^{-5}$ times the size of the 
interlevel separation energy. 
Several observations can be made: 

1) In all cases shown, the ``upper'' bands (which happen 
to be electronically 
excited) exhibit (Figs 3, 4, 7-9)
$\Delta I=2$ staggering which is 2 to 3 orders of magnitude 
larger than the experimental error, while 
the corresponding ``lower'' bands (which, in the cases studied, correspond to
the electronic ground state of each molecule), show (Figs 5, 6)
some effect smaller than the experimental error. 

2) There is no uniform dependence of the $\Delta I=2$ staggering on the 
angular momentum $I$. In some cases of long bands, though, it appears that 
the pattern is a sequence of points exhibiting small staggering, 
interrupted by groups of 6 points each time showing large staggering. 
The best examples can be seen in Figs  3a, 3b, 7a, 7b. In Fig. 3a
(odd levels of the $v=1$ C$^1 \Sigma ^+$ band of YD)) 
the first group of points showing appreciable $\Delta I =2$ 
staggering appears at $I=13$--23, while the second group appears at
$I=27$--37. In Fig. 3b (even levels of the $v=1$ C$^1 \Sigma^+$ band of YD) 
the first group appears at $I=12$--22, 
while the second group at $I=26$--36. In Fig. 7a (odd levels of the $v=0$ 
A$^6 \Sigma ^+$ band of CrD)
the first group
appears at $I=15$--25, while the second at $I=27$--37. Similarly 
in Fig. 7b (even levels of the $v=0$ A$^6 \Sigma ^+$ band of CrD) 
the first group appears at $I=14$--24, while the second 
group at $I=26$--36. 

3) In all cases shown, the results obtained for the odd levels of a band
are in good agreement with the results obtained for the even levels of the 
same band. For example, the regions showing appreciable staggering 
are approximately the same in both cases (compare Fig. 3a with Fig. 3b
and Fig. 7a with Fig. 7b, already discussed in 2)~).  In addition, the 
positions of the local staggering maxima in each pair of figures are 
closely related. In Fig. 3a, for example, maximum staggering appears at
$I=19$ and $I=31$, while in Fig. 3b the maxima appear at $I=18$ and 
$I=32$. 

4) In several cases the $\Delta I=2$ staggering of a band can be calculated 
from two different sets of data. For example, Figs 3a, 3b show the 
$\Delta I=2$ staggering of the $v=1$ C$^1 \Sigma^+$ band of YD calculated 
from the data on the 1--1 C$^1 \Sigma ^+$--X$^1\Sigma^+$ transitions,
while Figs 3c, 3d show the staggering of the same band calculated from the 
data on the 1--2 C$^1 \Sigma ^+$--X$^1 \Sigma^+$ transition. 
We remark that the results 
concerning points showing staggering larger than the experimental error 
come out completely consistently from the two calculations (region 
with $I=13$--23 in Figs 3a, 3c; region with $I=12$--22 in Figs 3b, 3d), 
while the results concerning points exhibiting staggering of the order 
of the experimental error come out randomly (in Fig. 3a, for example,
$I=11$ corresponds to a local minimum, while in Fig. 3c it corresponds to
a local maximum). Similar results are seen in the pairs of figures 
(3b, 3d), (4a, 4c), (4b, 4d), (6a, 6c), (6b, 6d), (9a, 9c), (9b, 9d).  
The best example of disagreement 
between staggering pictures of the 
same band calculated from two different sets of data is offered 
by Figs 6b, 6d, which concern the $v=2$ X$^1 \Sigma ^+$ band of YD,
which shows staggering of the order of the experimental error.

5) When considering levels of the same band, in some cases the odd levels 
exhibit larger staggering than the even levels, while in other cases the 
opposite is true. In the $v=1$ C$^1\Sigma^+$ band of YD, for example, 
the odd levels (shown in Fig. 3a, corroborated by Fig. 3c) show staggering
larger than that of the even levels (shown in Fig. 3b, corroborated by
Fig. 3d), while in the $v=2$ C$^1\Sigma^+$ band of YD the odd levels 
(shown in Fig. 4a, corroborated by Fig. 4c) exhibit staggering smaller 
than that of the even levels (shown in Fig. 4b, corroborated by Fig. 4d). 

\bigskip
{\bf IV. Discussion}
\medskip

The observations made above can be explained by the assumption that 
the staggering observed is due to the presence of one or more bandcrossings
\cite{Pavli,MRM}.
The following points support this assumption:

1) It is known \cite{VDS}  that bandcrossing occurs in cases in which 
the interband interaction is weak. In such cases only the one or two levels 
closest to the crossing point are affected \cite{Bonbb}. 
However, if one level is influenced 
by the crossing, in the corresponding staggering figure six points get 
influenced. For example, if E(16) is influenced by the crossing, 
the quantities $DE_2(15)$ and $DE_2(17)$ are influenced (see Eq. (10)~), 
so that in the corresponding figure the points $\Delta E_2(I)$ with 
$I=11$, 13, 15, 17, 19, 21 are influenced,  as seen from Eq. (11). 
This fact explains why points showing 
appreciable staggering appear in groups of 6 at a time. 

2) It is clear that if bandcrossing occurs, large staggering should appear 
in approximately 
the same angular momentum regions of both even levels and odd levels. 
As we have already seen, this is indeed the case. 

3) It is clear that when two bands cross each other, maximum staggering 
will appear at the angular momentum for which the energies of the relevant 
levels of each band are approximately equal \cite{Bonbb}. 
If this angular momentum 
value happens to be odd, then $\Delta E_2(I)$ for even values of $I$ 
 in this region (the group
of 6 points centered at this $I$) will show larger staggering than the 
$\Delta E_2(I)$ for odd values of $I$ in the corresponding region, 
and vice versa. For example, 
if the closest approach of two bands occurs for $I=31$, then $\Delta E_2(I)$ 
for even values of $I$ in the $I=26$--36 region will show larger staggering 
than $\Delta E_2(I)$ for odd values of $I$ in the same region. This is in 
agreement with the empirical 
observation that in some cases the odd levels show larger staggering than
the even levels, while in other cases the opposite holds. 

4) The presence of staggering in the ``upper'' (electronically excited)
bands and the lack of staggering in the ``lower'' (electronic ground 
state) bands can be attributed to the fact that the electronically 
excited bands have several neighbours with 
which they can interact, while the bands built on the electronic 	
ground state are relatively isolated, and therefore no bandcrossings 
occur in this case. In the case of the CrD molecule, in particular, 
it is known \cite{CrD} that there are many strong Cr atomic lines 
present, which frequently overlap the relatively weaker 
(electronically excited) molecular lines. In addition, Ne atomic lines 
are present \cite{CrD}. Similarly, in the case of the 
YD molecule the observed spectra are influenced by Y and Ne atomic lines
\cite{YD}, while in the case of the CrH molecule there are Ne and Cr 
atomic lines influencing the molecular spectra \cite{CrH}. 

5) The fact that consistency between results for the same band calculated 
from two different sets of data is observed only in the cases in which
the staggering is much larger than the experimental error, corroborates 
the bandcrossing explanation. The fact that the results obtained in areas
in which the staggering is of the order of the experimantal error, or 
even smaller, appear to be random, points towards the absence of any 
real effect in these regions. 

It should be noticed that bandcrossing has been proposed \cite{RJR,HS,HL}
as a possible explanation for the appearance of $\Delta I=2$ staggering 
effects in normally deformed nuclear bands \cite{Wu,RJR,HL} and 
superdeformed nuclear bands \cite{HS}. 

The presence of two subsequent bandcrossings can also provide an explanation 
for the effect of mid-band disappearance of $\Delta I=2$ staggering 
observed in superdeformed bands of some Ce isotopes \cite{Semple}. 
The effect seen in the Ce isotopes is very similar to the mid-band 
disappearance of staggering seen, for example, in Fig. 3a. 

\bigskip
{\bf V. Conclusion}
\medskip

In conclusion, we have found several examples of $\Delta I=2$ staggering
in electronically excited bands of diatomic molecules. The details of 
the observed effect are in agreement with the assumption that it is due 
to one or more bandcrossings. In these cases the magnitude of the effect 
is clearly larger than the experimental error. In cases in which 
an effect of the order of the experimental error appears, we have shown 
that this is an artifact of the method used, since different sets of data 
from the same experiment and for the same molecule lead to different 
staggering results for the same rotational band. The present work 
emphasizes the need to ensure in all cases (including staggering candidates
in nuclear physics) that the effect is larger than the experimental 
error and, in order to make assumptions about any new symmetry,
that it is not due to a series of bandcrossings.

\bigskip
{\bf Acknowledgements}
\medskip

Two of the authors (DB,CD) acknowledge many helpful discussions with 
Professor John F. Ogilvie of the Oregon State University.  
One of the authors (PPR) acknowledges support from the Bulgarian Ministry 
of Science and Education under contracts $\Phi$-415 and $\Phi$-547, 
while another author (NM) acknowledges support from the Bulgarian National 
Fund for Scientific Research under contract MU-F-02/98.    
Three authors (DB,CD,NK) have been supported by the Greek Secretariat 
of Research and Technology under contract PENED 95/1981.

\newpage

\centerline{\bf Figure captions}

\begin{itemize}

\item[{\bf Fig. 1}] $\Delta E_2(I)$ (in keV), calculated  from Eq. (2), 
versus the angular frequency $\omega$ (in MeV), calculated from Eq. (7), 
for various superdeformed bands in the nucleus $^{149}$Gd \cite{Fli}. 
a) Band (a) of Ref. \cite{Fli}. b) Band (d) of Ref. \cite{Fli}. 

\item[{\bf Fig. 2}] $\Delta E_2(I)$ (in keV), calculated from Eq. (2), 
versus the angular frequency $\omega$ (in MeV), calculated from Eq. (7),
for various superdeformed bands in the nucleus $^{194}$Hg \cite{Ced}. 
a) Band 1 of Ref. \cite{Ced}. b) Band 2 of Ref. \cite{Ced}. 

\item[{\bf Fig. 3}] $\Delta E_2(I)$ (in cm$^{-1}$), calculated from Eq.
(11),  
for various bands of the YD molecule
\cite{YD}. 
a) Odd levels of the $v=1$ C$^1\Sigma^+$ band calculated from the 
data of the 1--1 C$^1\Sigma^+$--X$^1\Sigma^+$ transitions. 
b) Even levels of the previous band. 
c) Odd levels of the $v=1$ C$^1\Sigma^+$ band calculated from the 1--2
C$^1\Sigma^+$--X$^1\Sigma^+$ transitions. 
d) Even levels of the previous band. 

\item[{\bf Fig. 4}] $\Delta E_2(I)$ (in cm$^{-1}$), calculated from Eq. (11), 
for various bands of the YD molecule
\cite{YD}. 
a) Odd levels of the $v=2$ C$^1\Sigma^+$ band calculated from the 
data of the 2--2 C$^1\Sigma^+$--X$^1\Sigma^+$ transitions. 
b) Even levels of the previous band. 
c) Odd levels of the $v=2$ C$^1\Sigma^+$ band calculated from the 2--3
C$^1\Sigma^+$--X$^1\Sigma^+$ transitions. 
d) Even levels of the previous band. 
The experimental error in all cases is $\pm 0.006$ cm$^{-1}$ and therefore
is hardly or not seen. 

\item[{\bf Fig. 5}] $\Delta E_2(I)$ (in cm$^{-1}$), calculated from Eq. (11),
for various bands of the YD molecule
\cite{YD}. 
a) Odd levels of the $v=1$ X$^1\Sigma^+$ band calculated from the 
data of the 1--1 C$^1\Sigma^+$--X$^1\Sigma^+$ transitions. 
b) Even levels of the previous band. 

\item[{\bf Fig. 6}] $\Delta E_2(I)$ (in cm$^{-1}$), calculated from Eq. (11), 
for various bands of the YD molecule
\cite{YD}. 
a) Odd levels of the $v=2$ X$^1\Sigma^+$ band calculated from the 
data of the 1--2 C$^1\Sigma^+$--X$^1\Sigma^+$ transitions. 
b) Even levels of the previous band. 
c) Odd levels of the $v=2$ X$^1\Sigma^+$ band calculated from the 2--2
C$^1\Sigma^+$--X$^1\Sigma^+$ transitions. 
d) Even levels of the previous band. 

\item[{\bf Fig. 7}] $\Delta E_2(I)$ (in cm$^{-1}$), calculated from Eq. (11), 
for various bands of the CrD molecule
\cite{CrD}. 
a) Odd levels of the $v=0$ A$^6\Sigma^+$ band calculated from the 
data (R2, P2 branches) of the 0--0 A$^6\Sigma^+$--X$^6\Sigma^+$ transitions. 
b) Even levels of the previous band. 
The experimental error in all cases is $\pm 0.006$ cm$^{-1}$
and therefore is not seen. 

\item[{\bf Fig. 8}] $\Delta E_2(I)$ (in cm$^{-1}$), calculated from Eq. (11), 
for various bands of the CrH molecule
\cite{CrH}. 
a) Odd levels of the $v=0$ A$^6\Sigma^+$ band calculated from the 
data (R2, P2 branches) 
of the 0--0 A$^6\Sigma^+$--X$^6\Sigma^+$ transitions. 
b) Even levels of the previous band. 
The experimental error in all cases is $\pm 0.004$ cm$^{-1}$
and therefore is not seen. 

\item[{\bf Fig. 9}] $\Delta E_2(I)$ (in cm$^{-1}$), calculated from Eq. (11), 
for various bands of the CoH molecule
\cite{CoH}. 
a) Odd levels of the $v=0$ A$'$$^{3}\Phi_4$ band calculated from the 
data (Ree, Pee branches) of the 0--1 A$'$$^{3}\Phi_4$--X$^3\Phi_4$ 
transitions. 
b) Even levels of the previous band. 
The experimental error in all cases is $\pm 0.01$ cm$^{-1}$ 
and therefore is not seen.

\end{itemize}

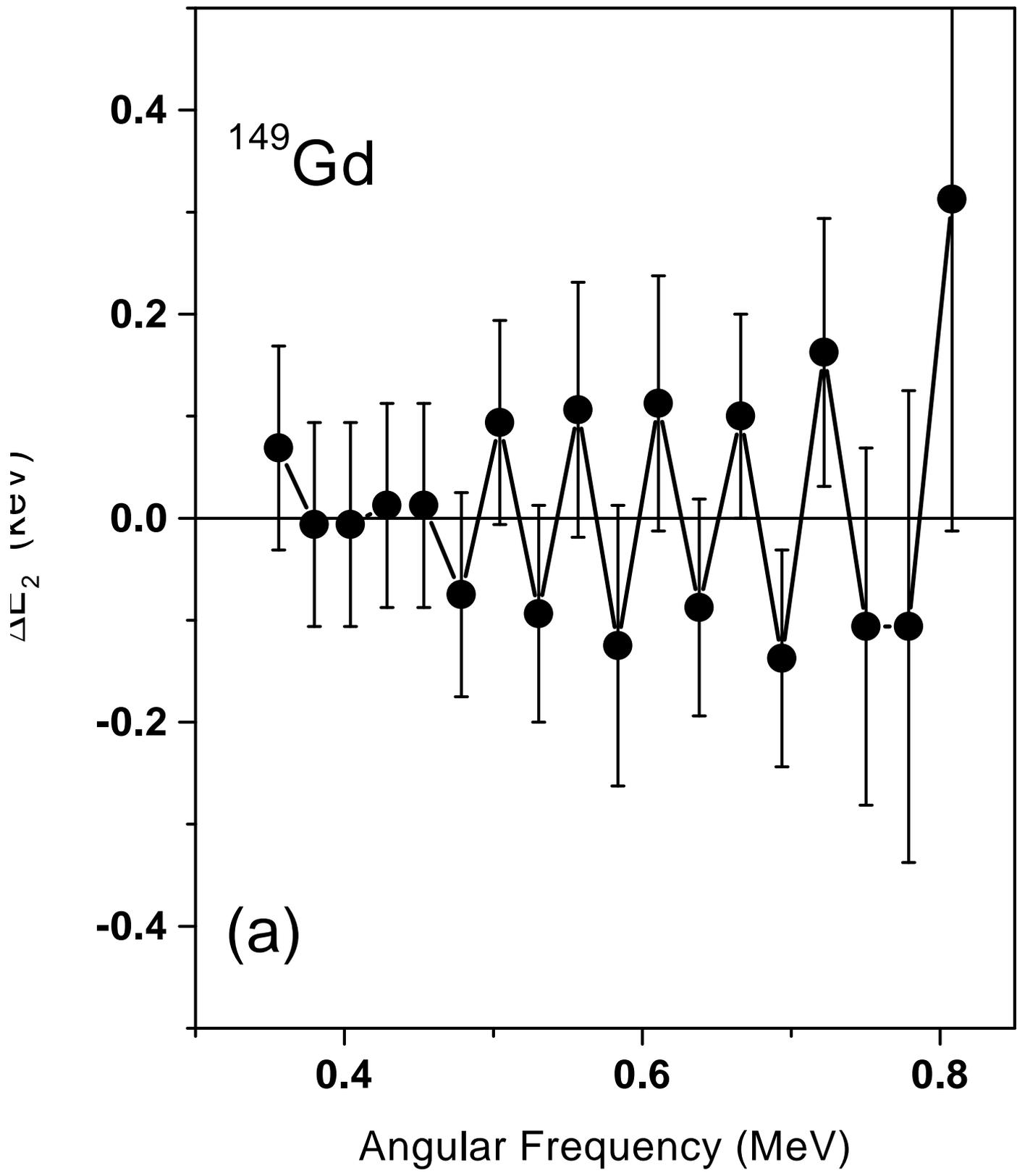
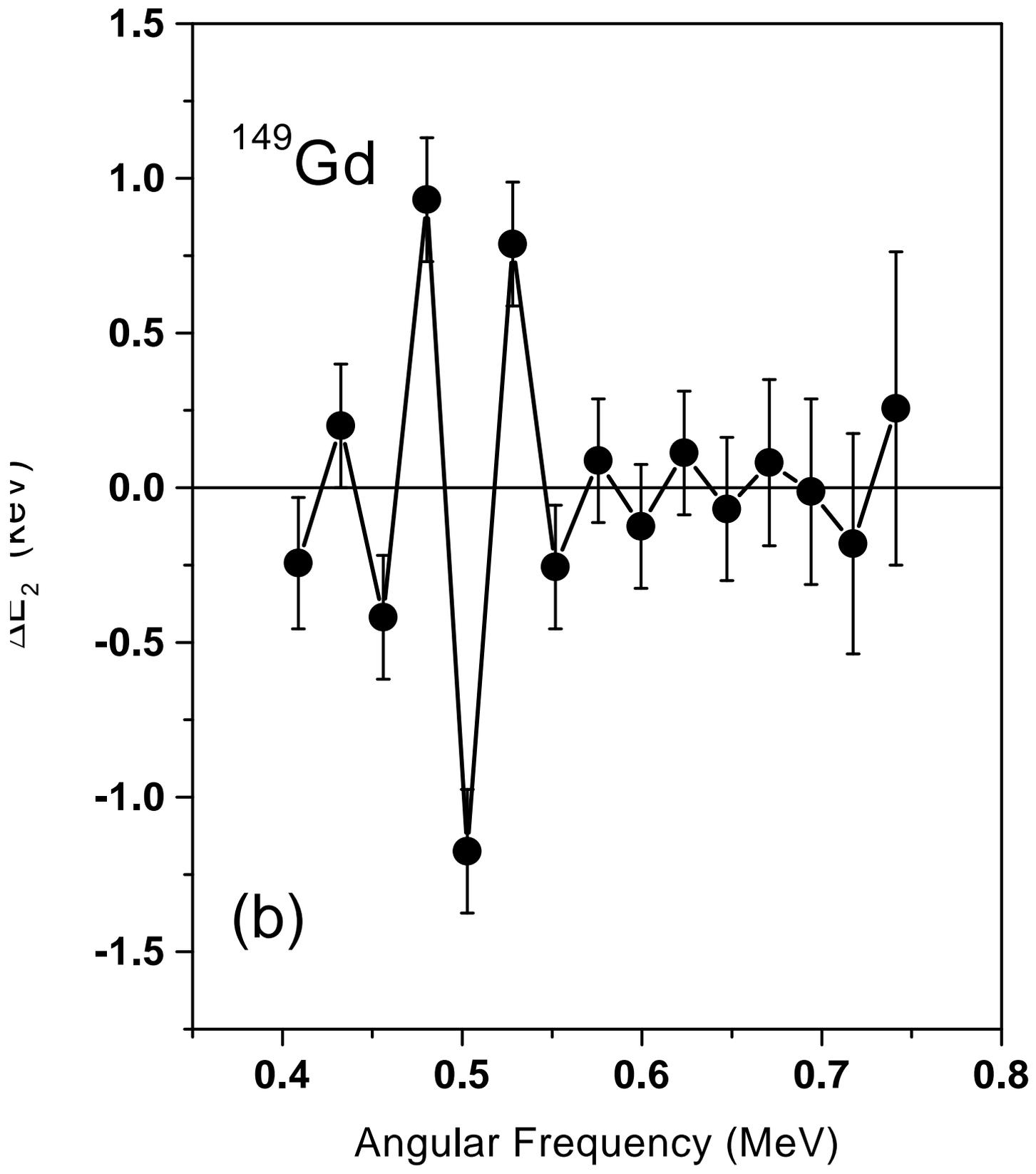
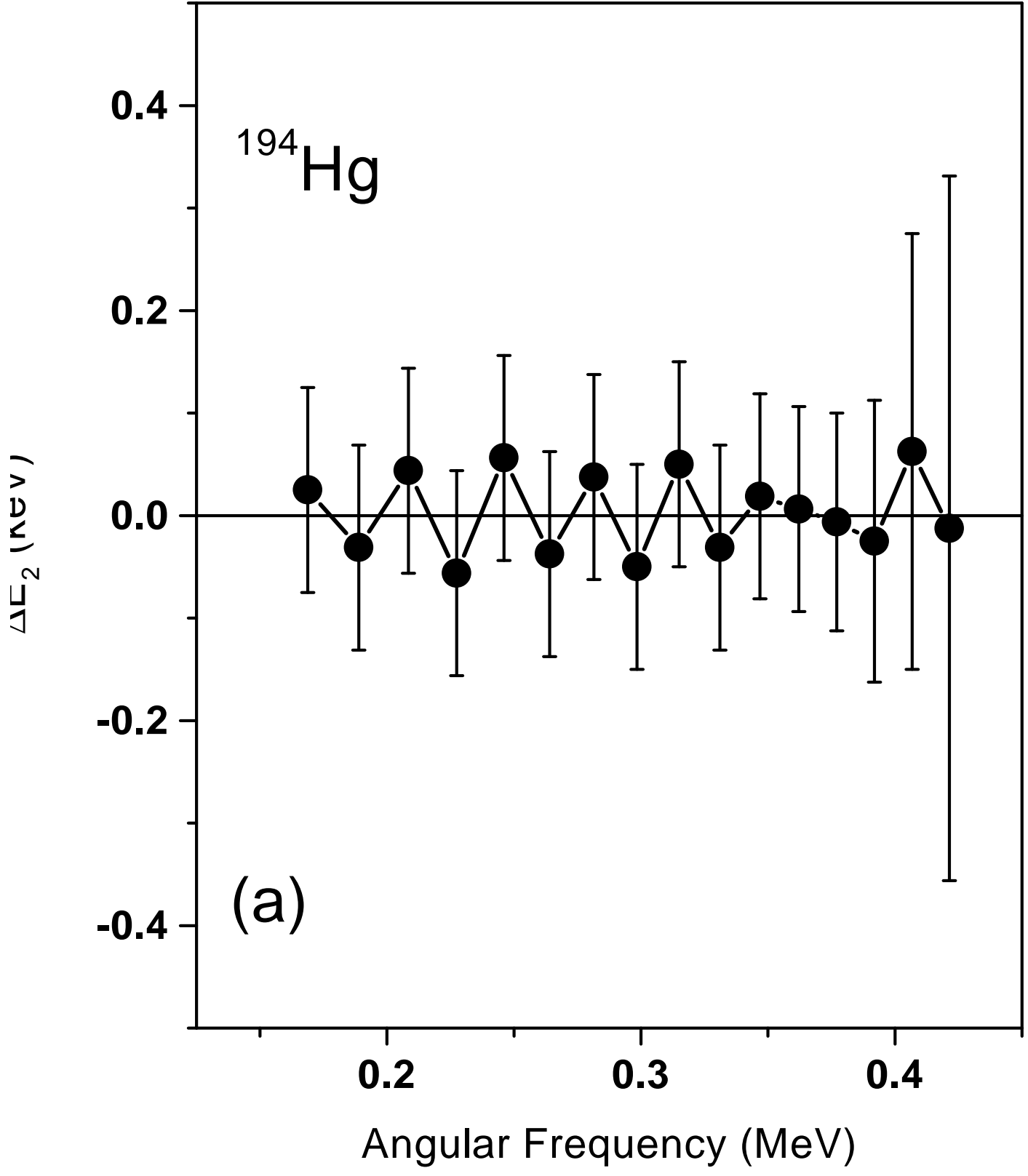
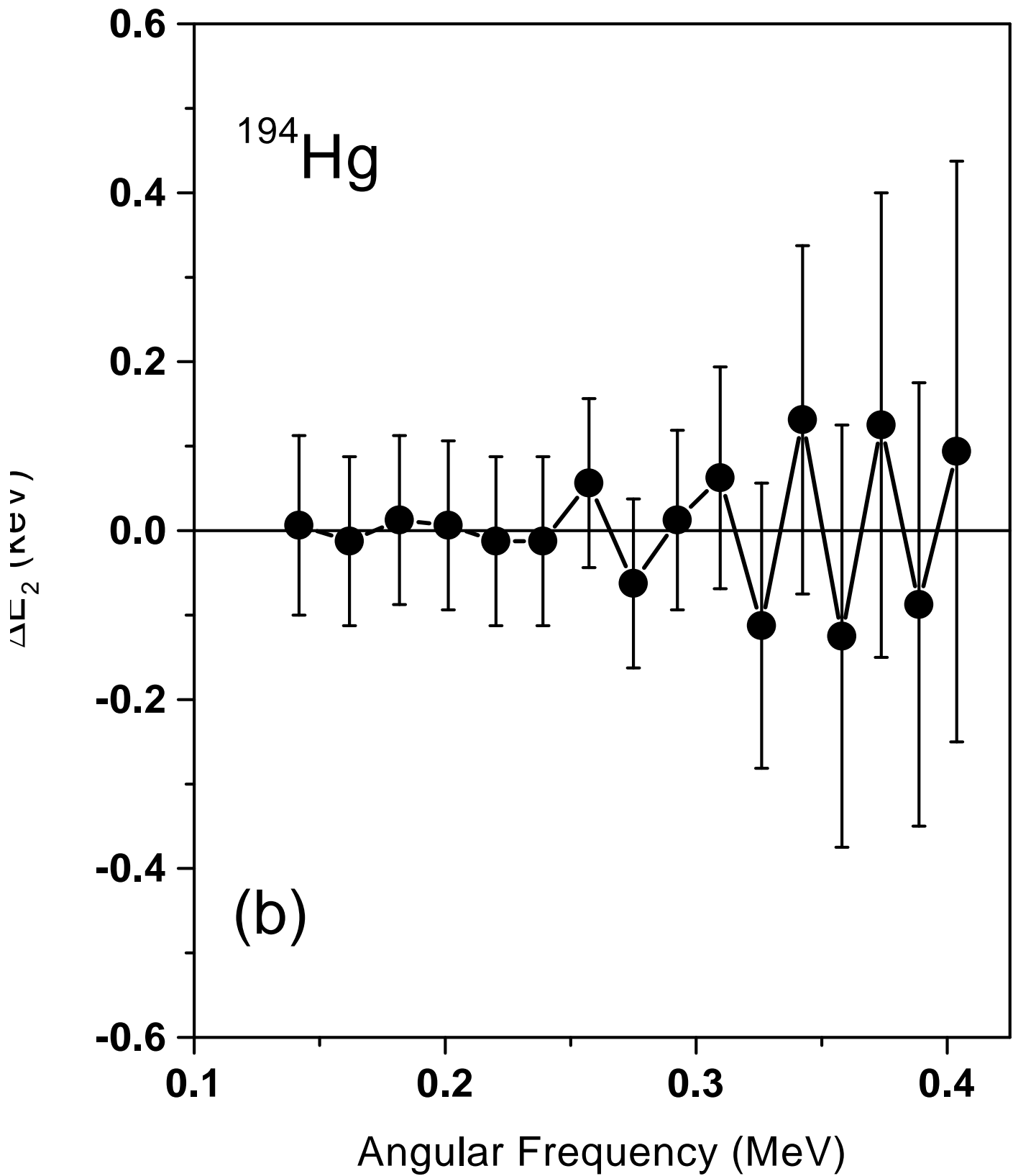
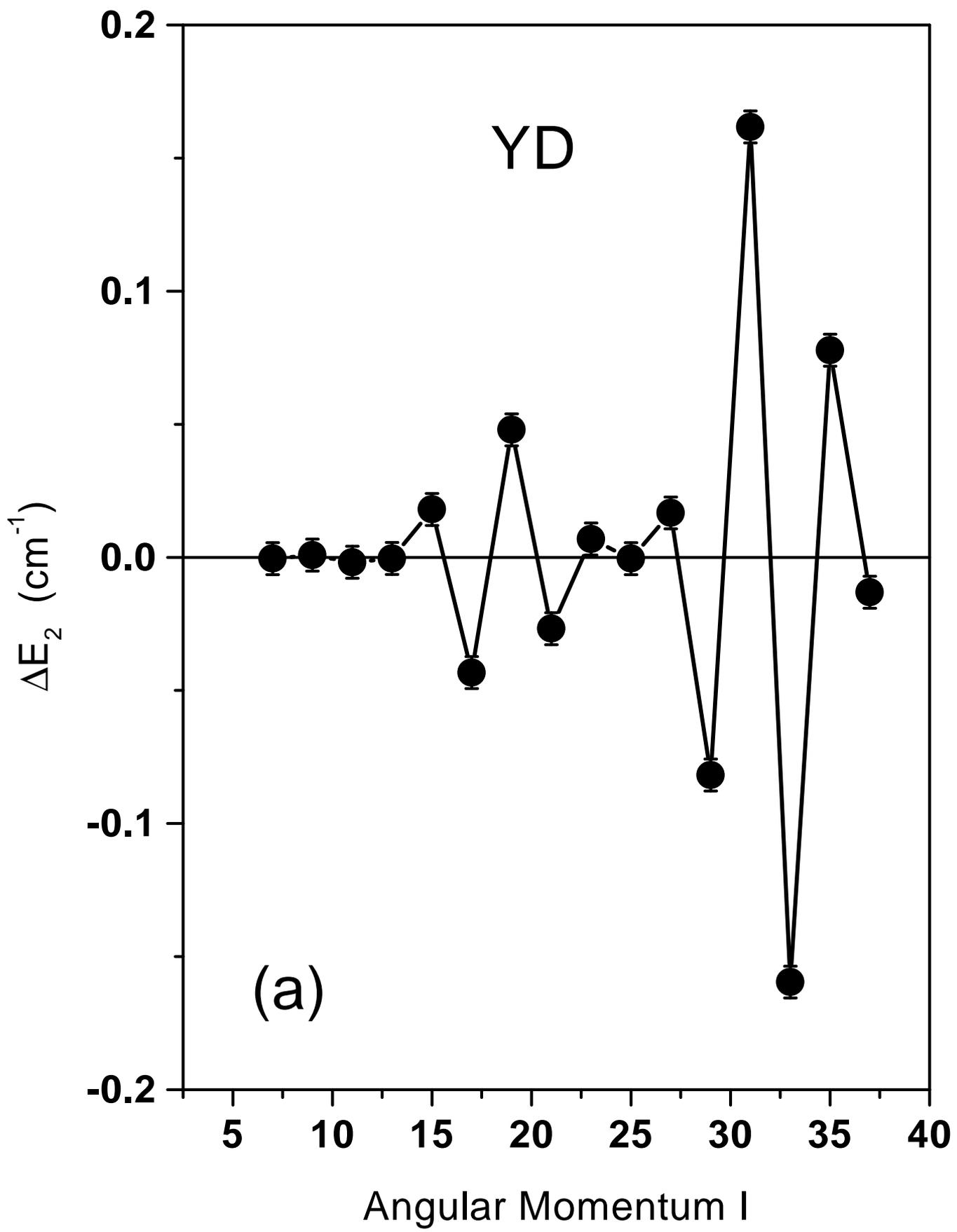
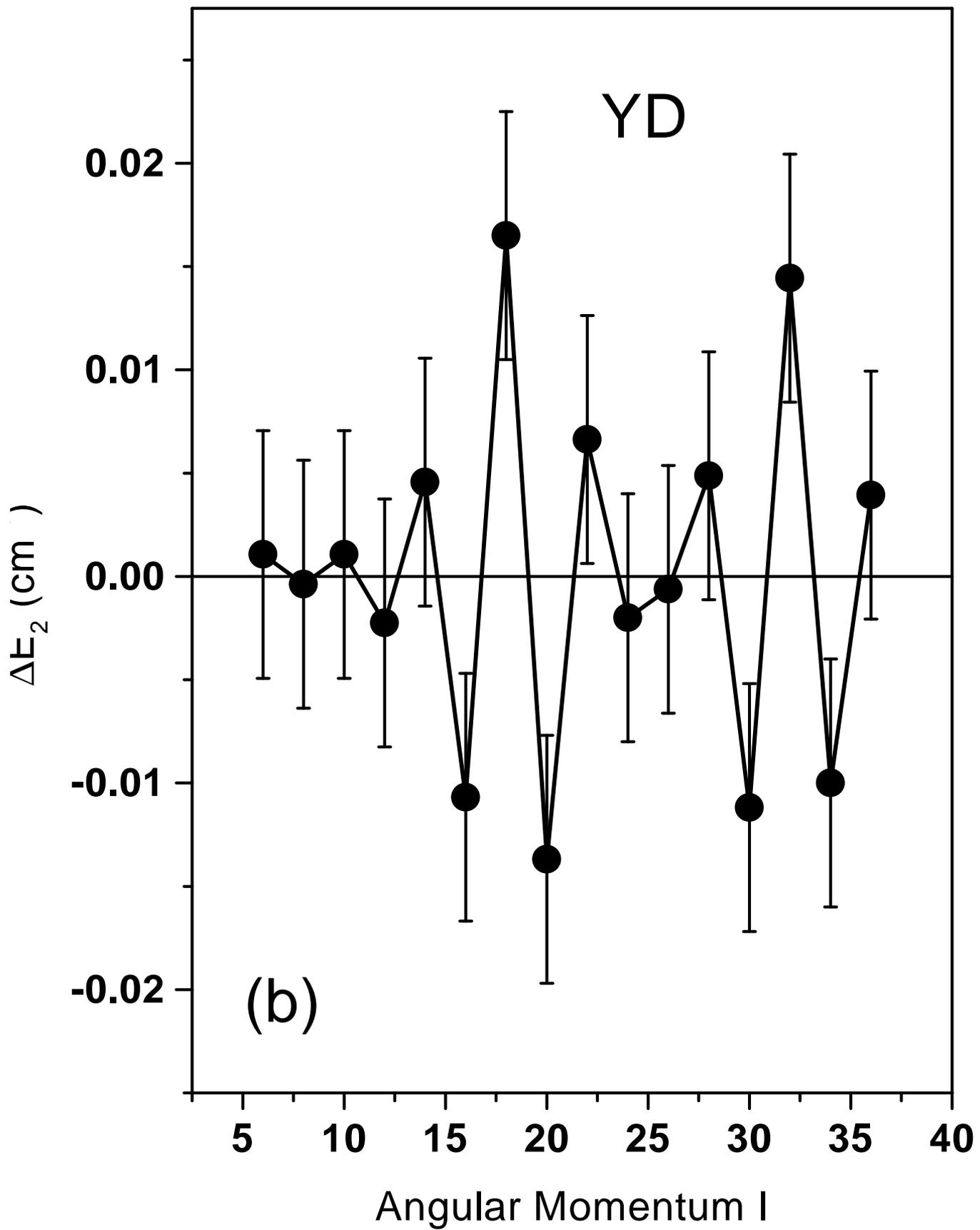
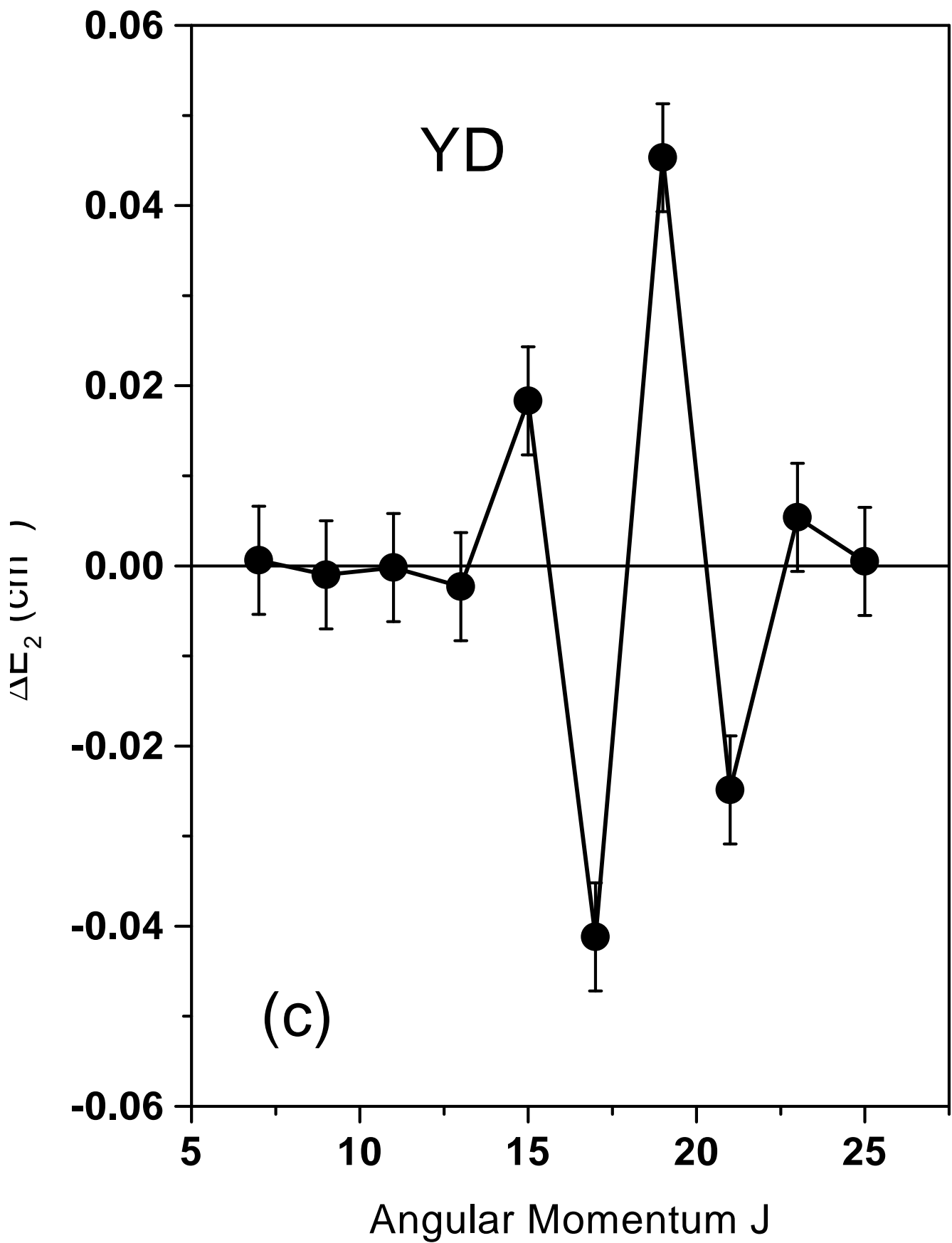
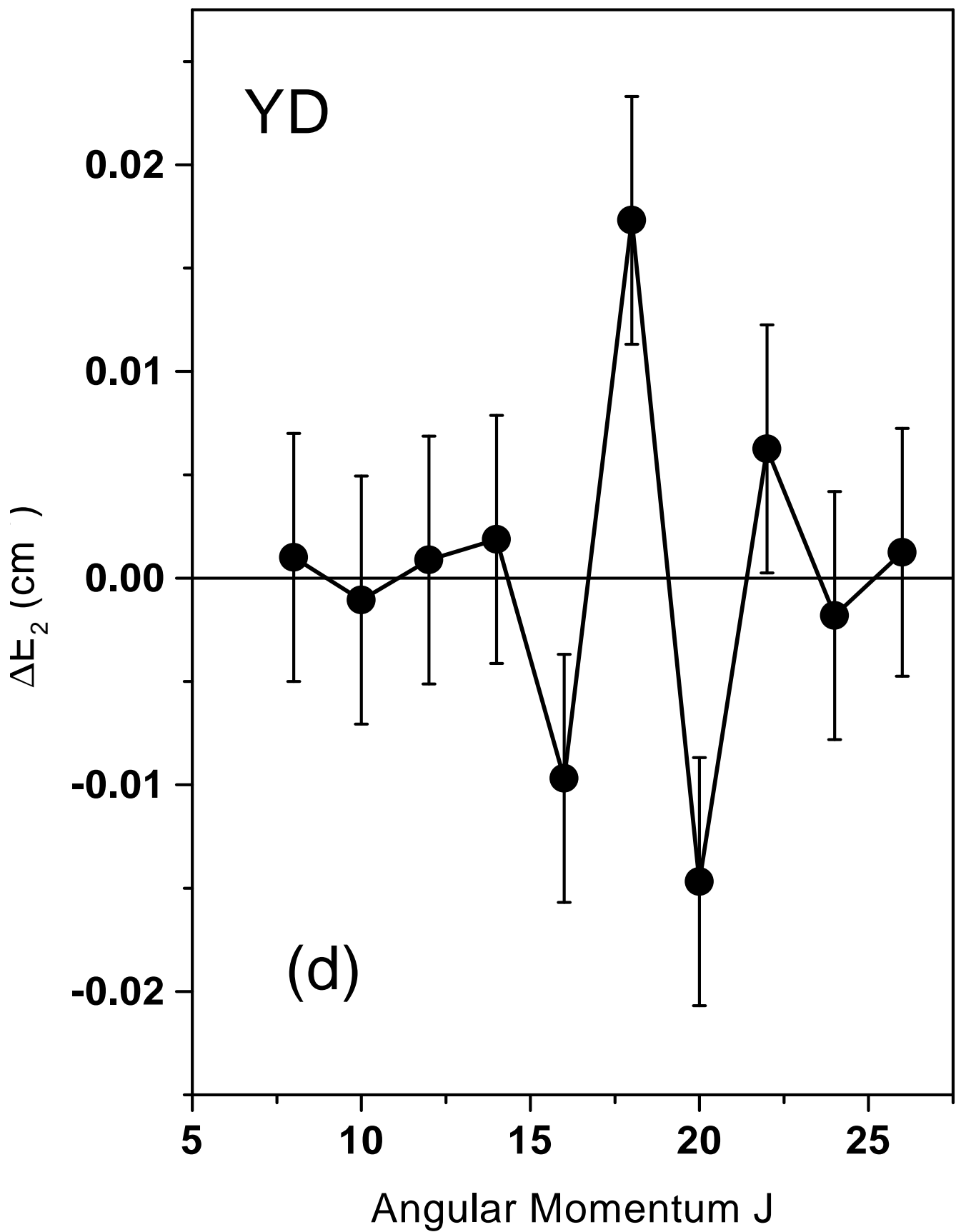
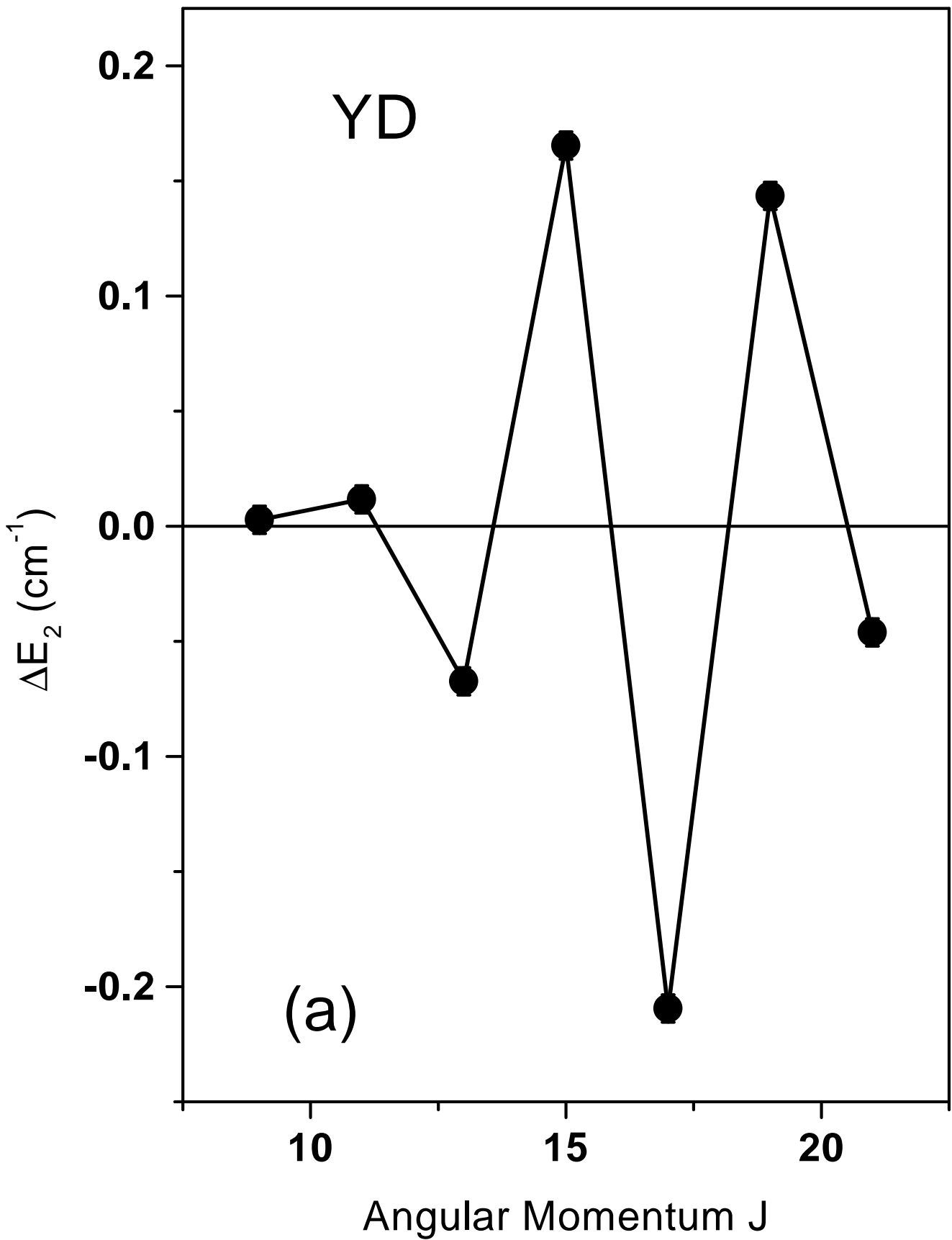
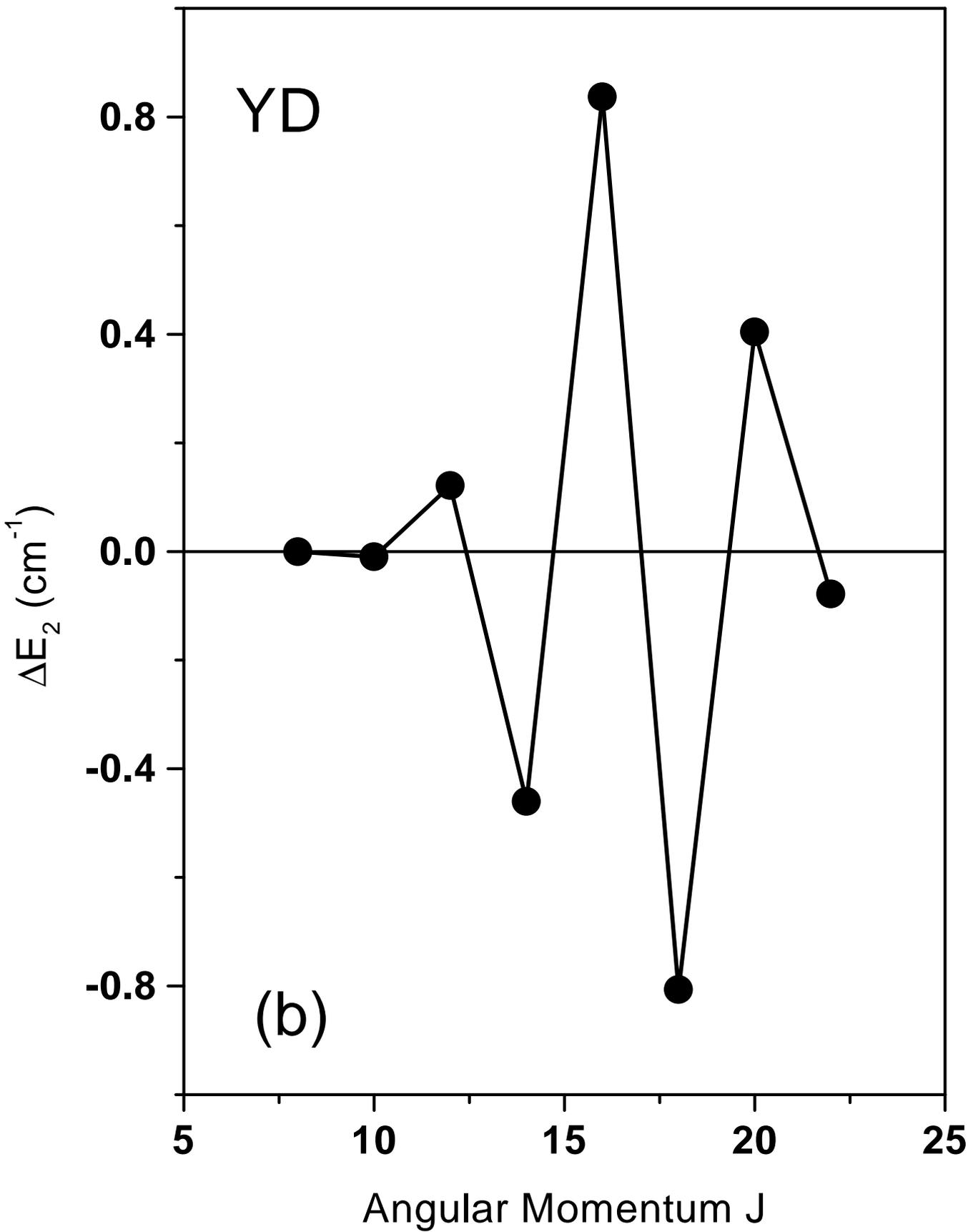
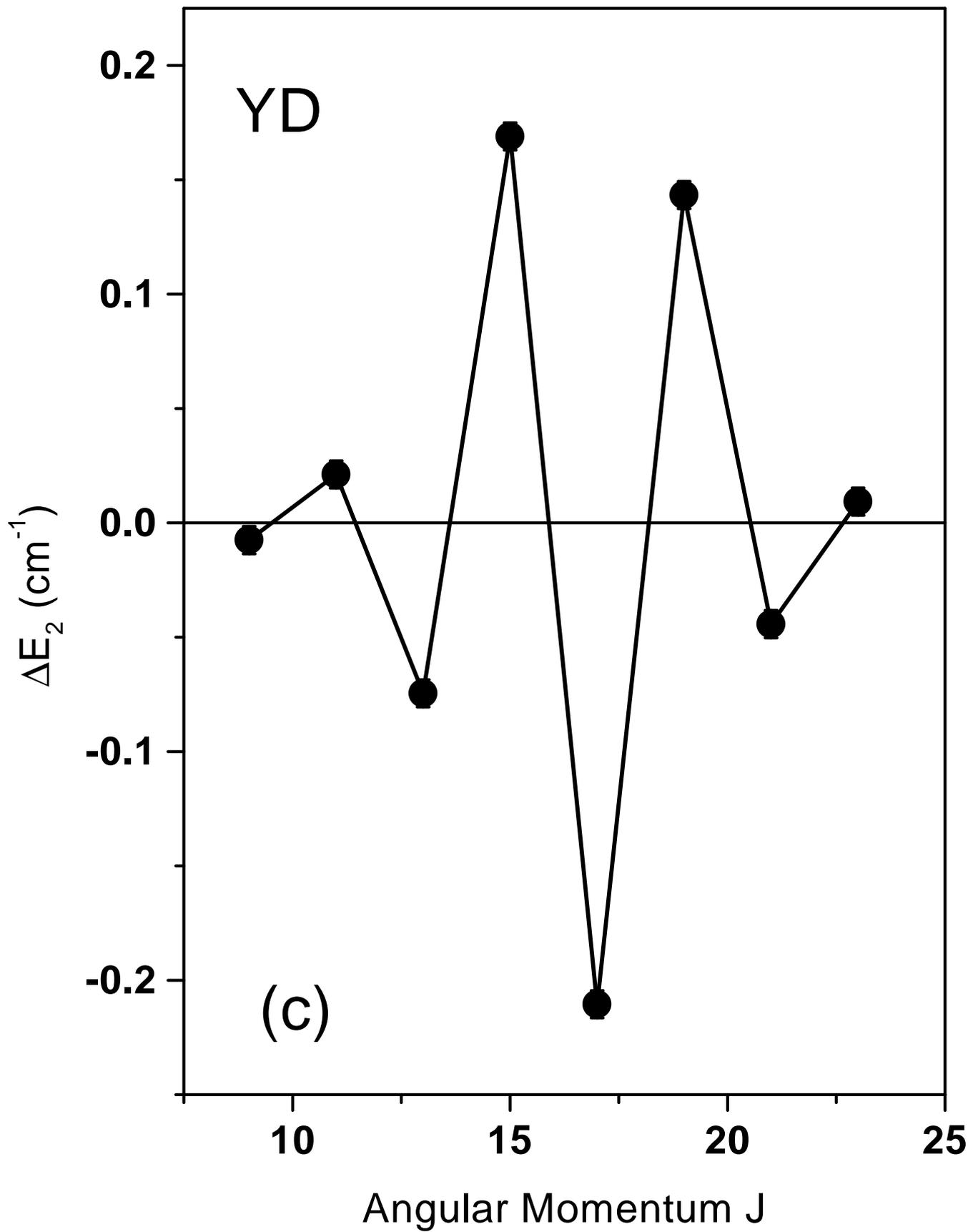
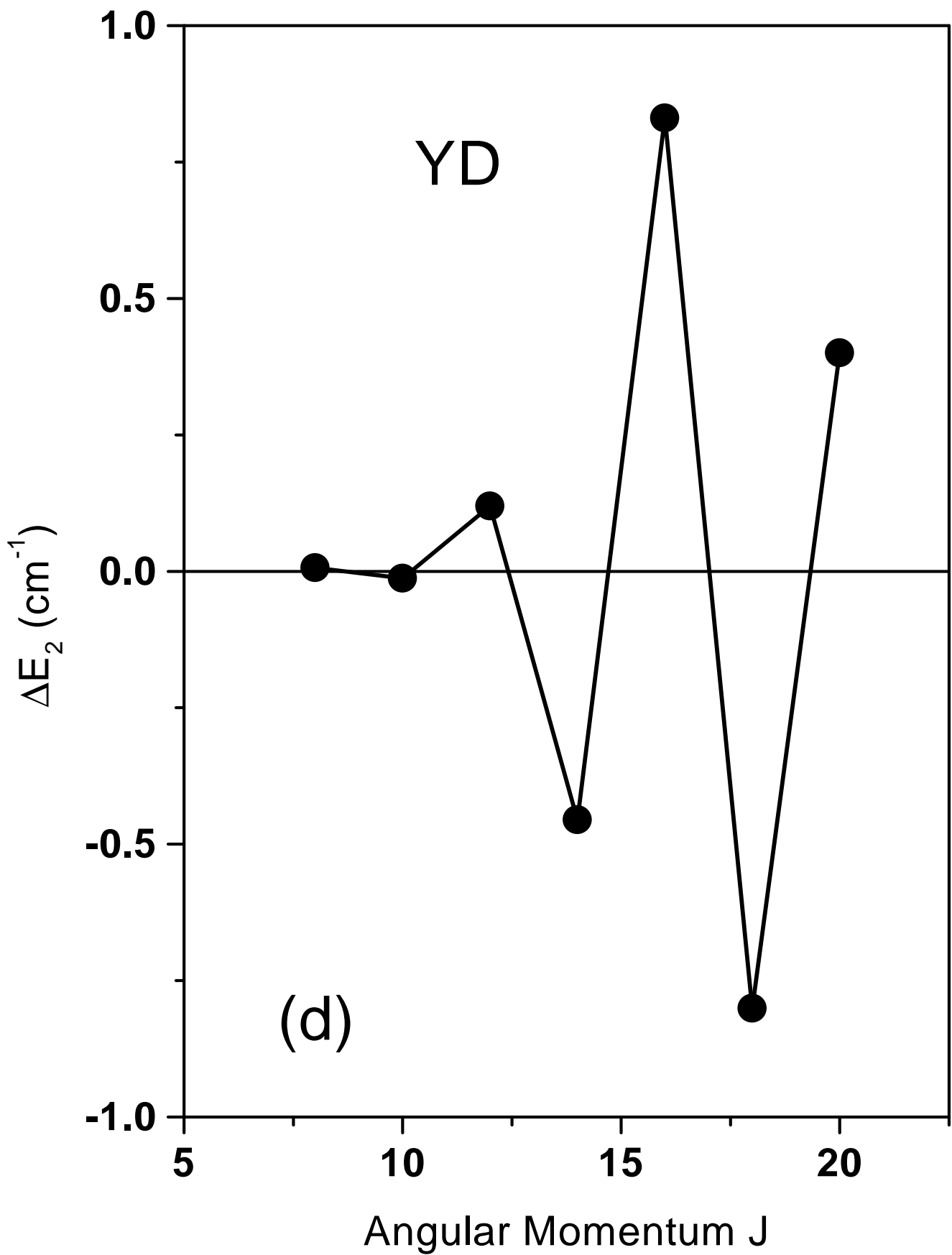
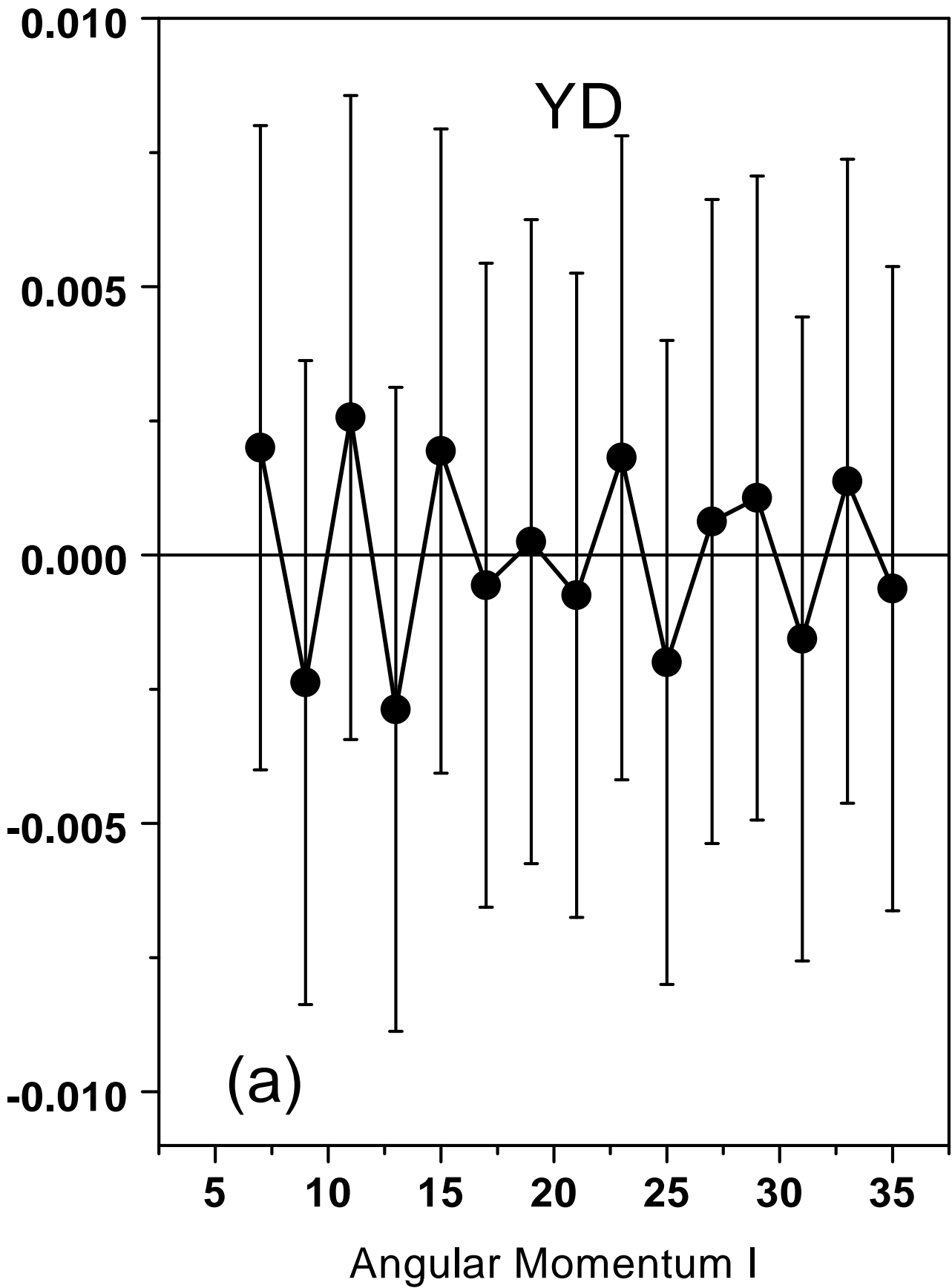
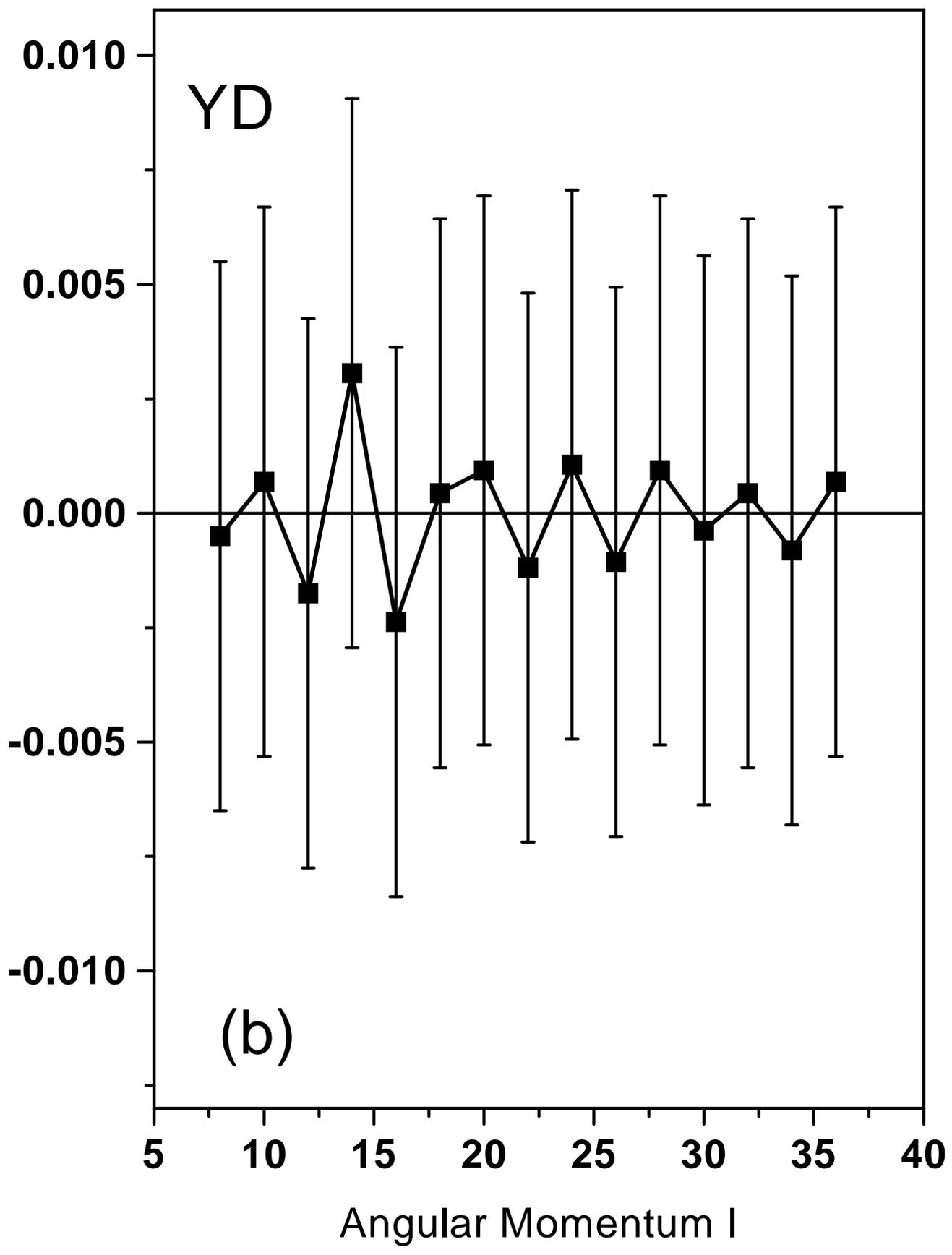
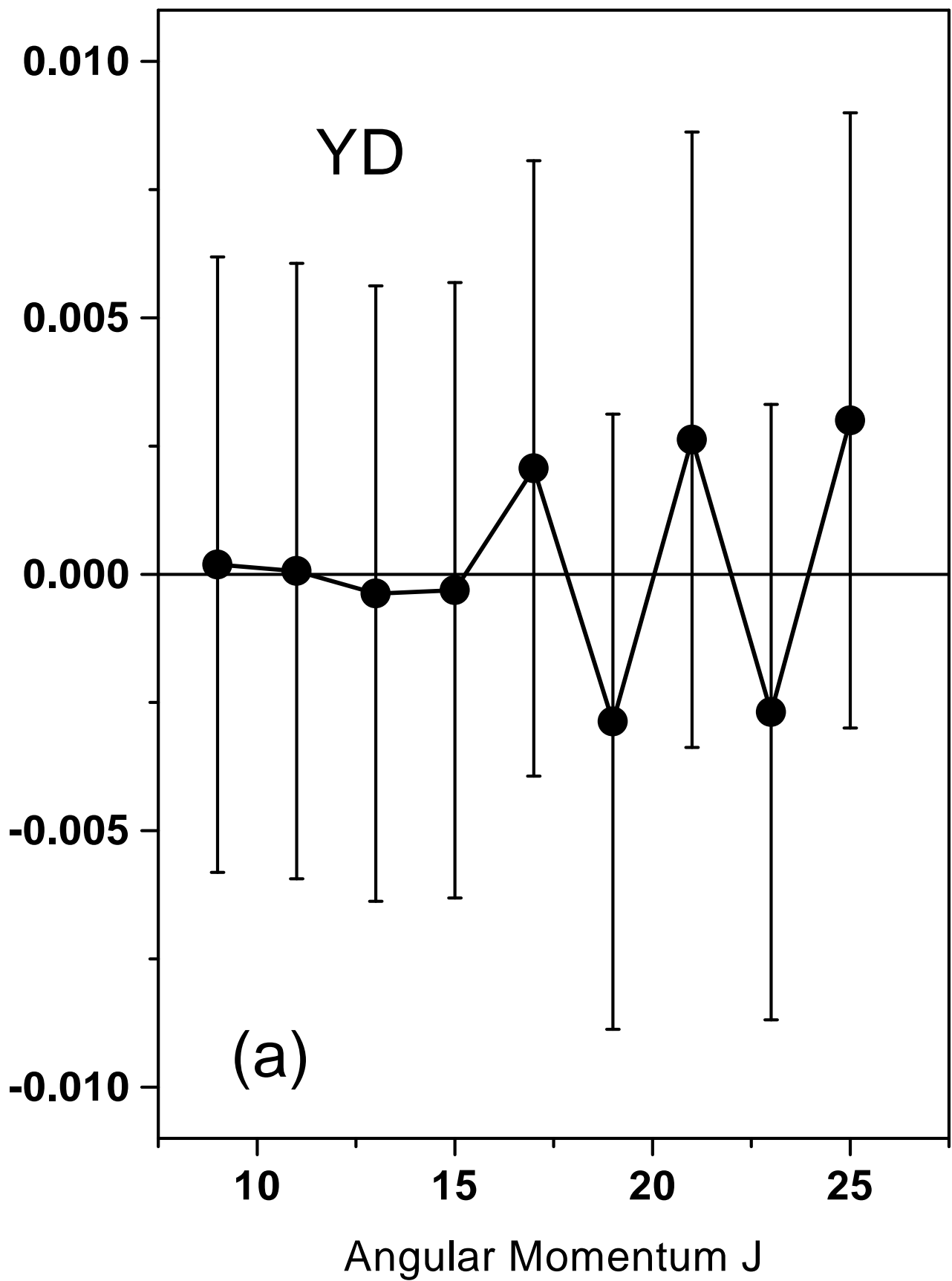
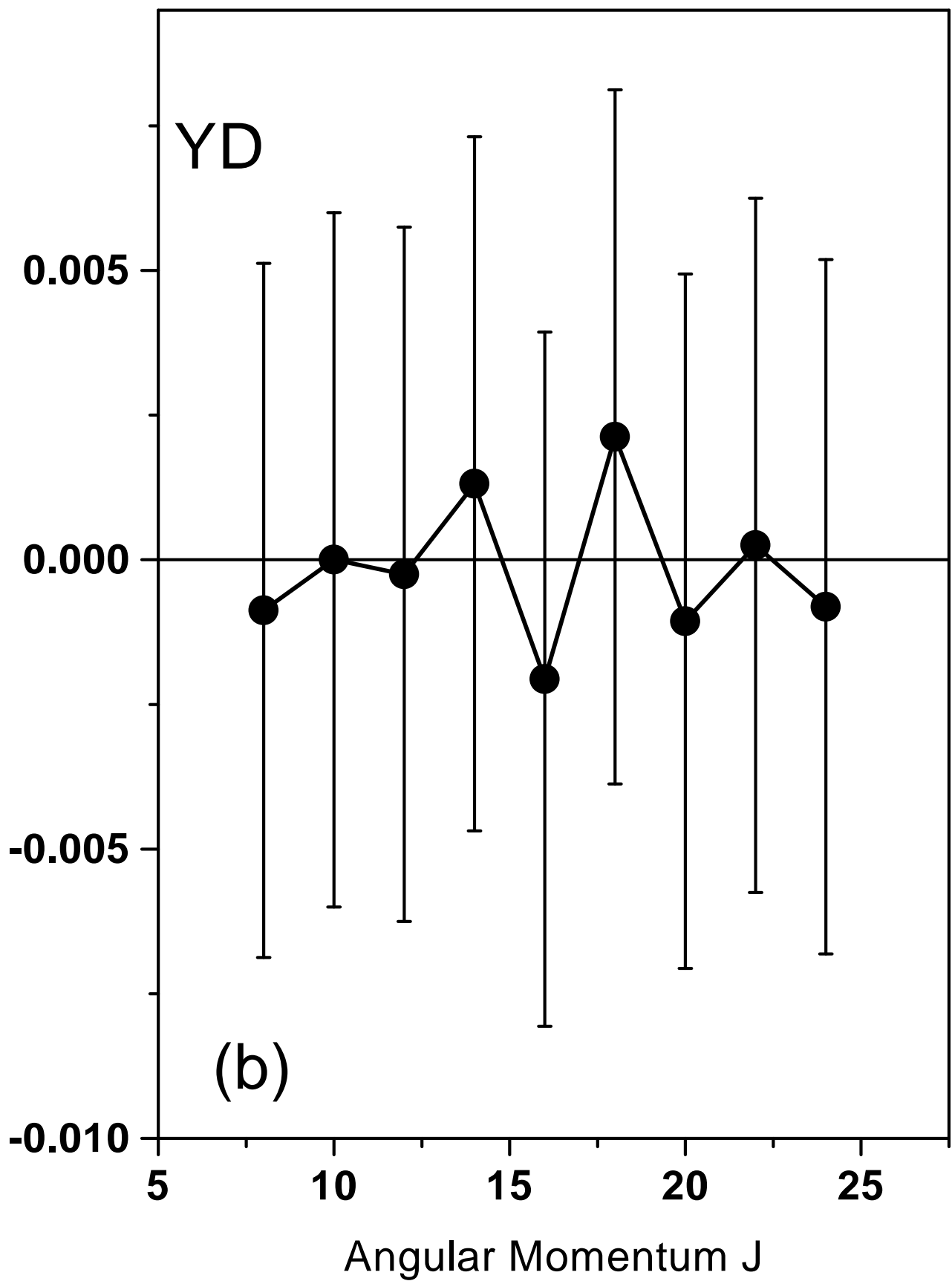
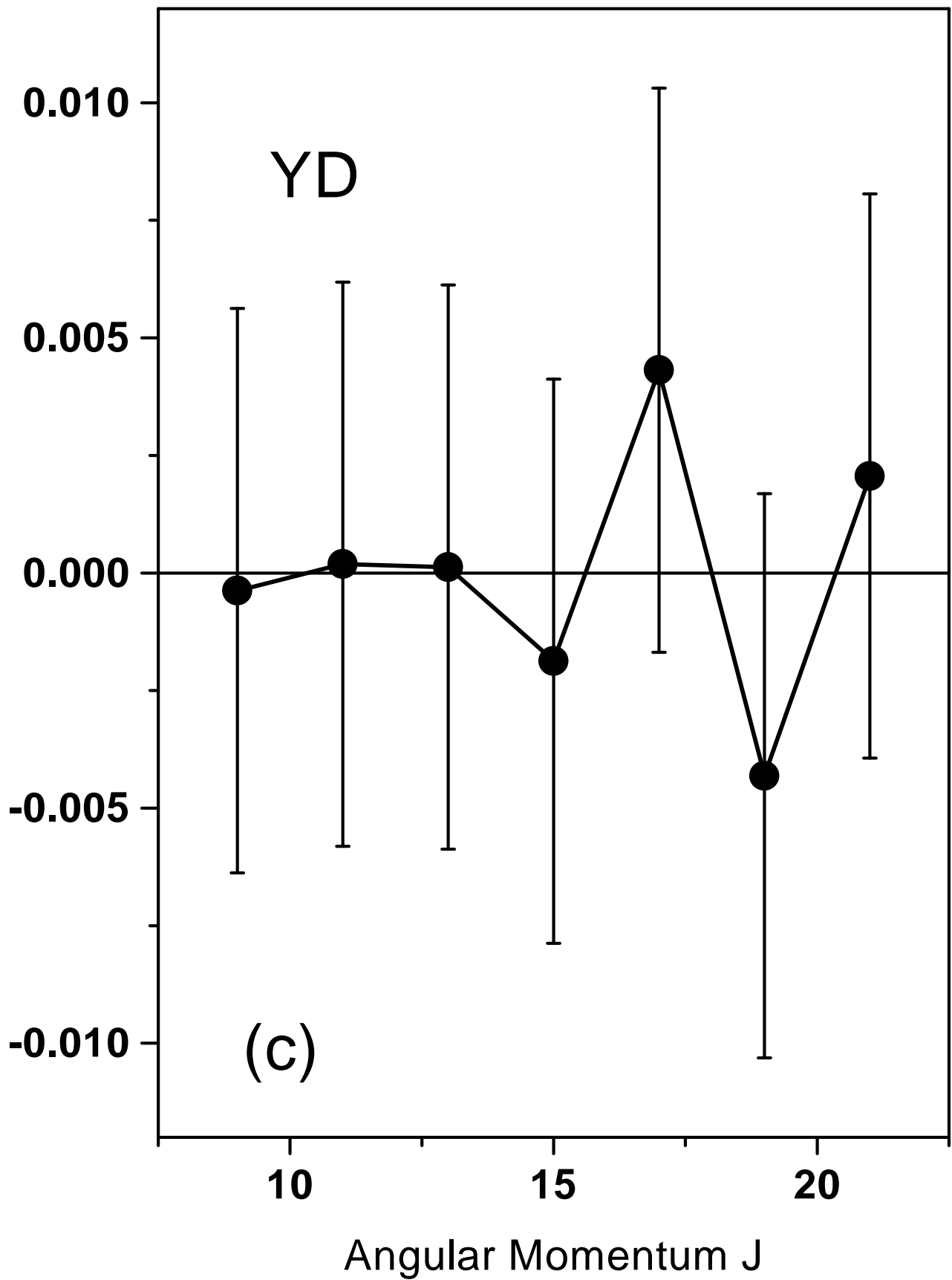
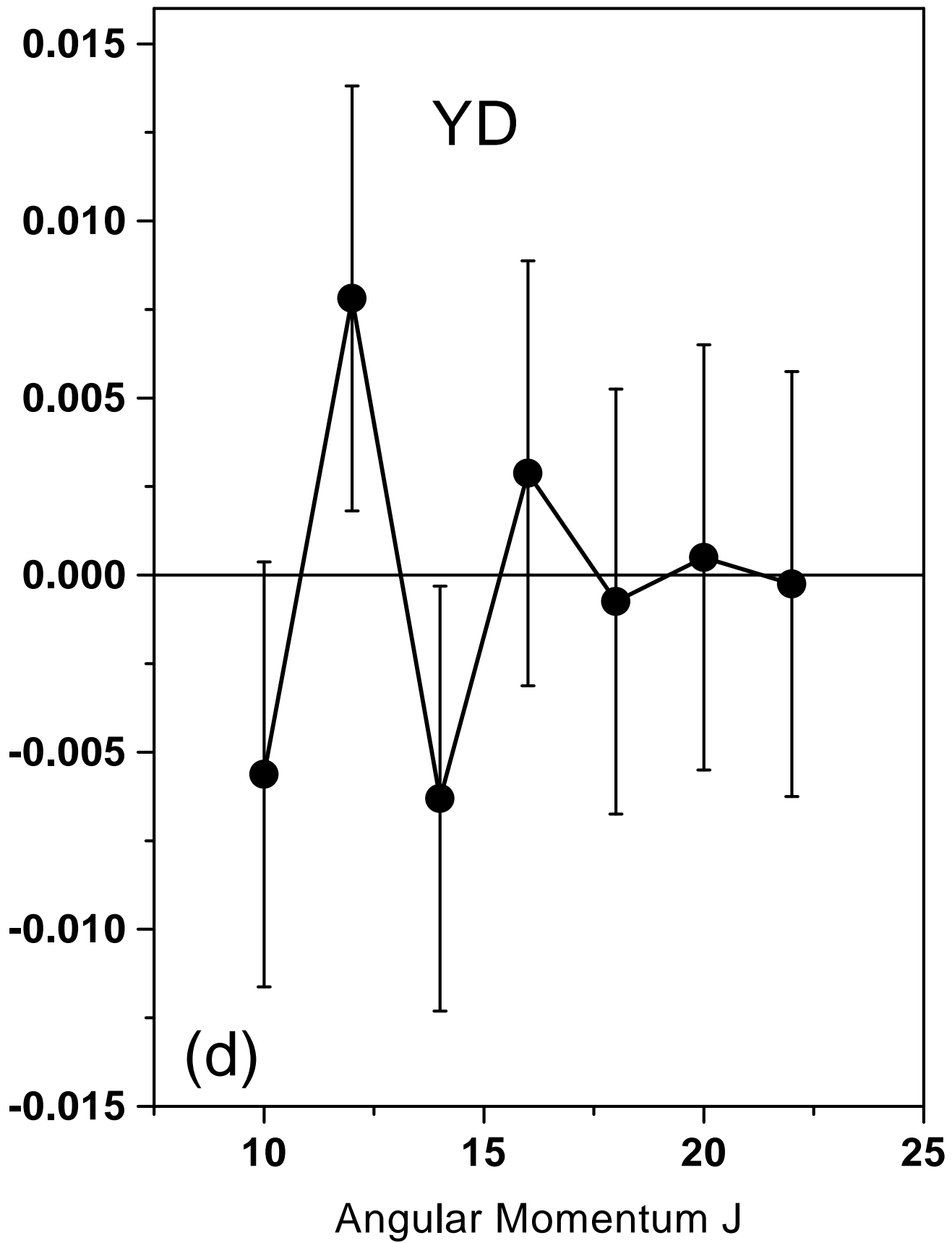
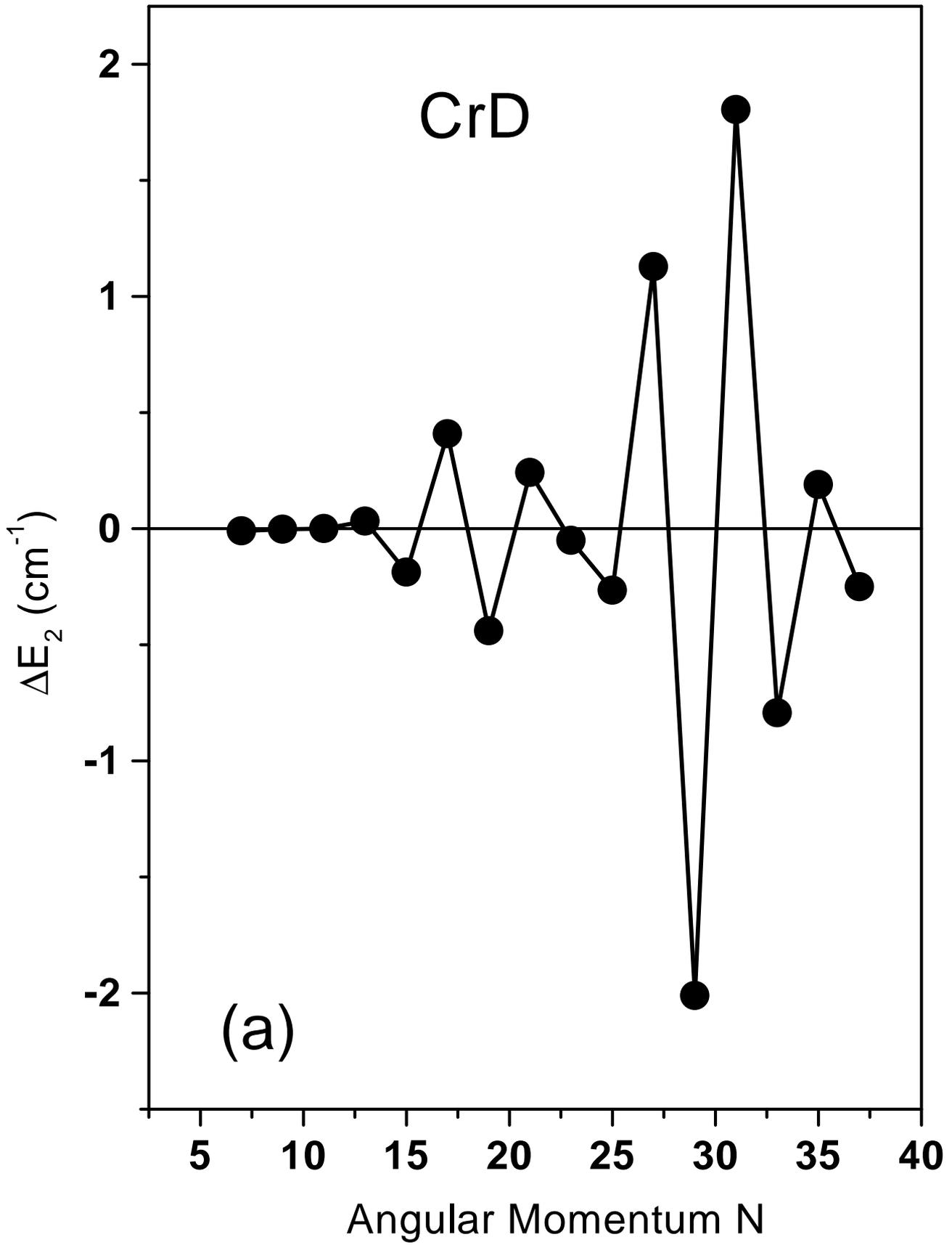
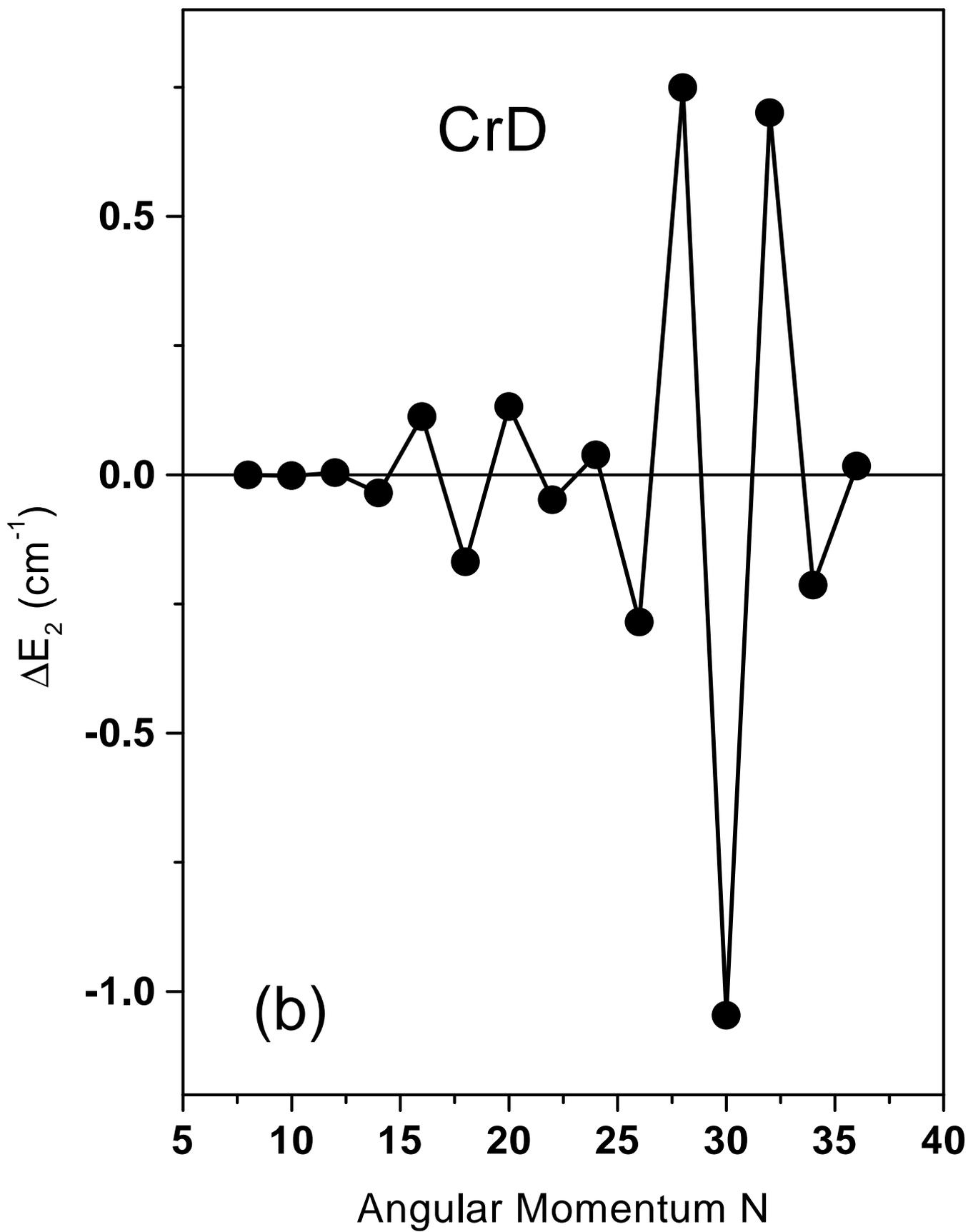
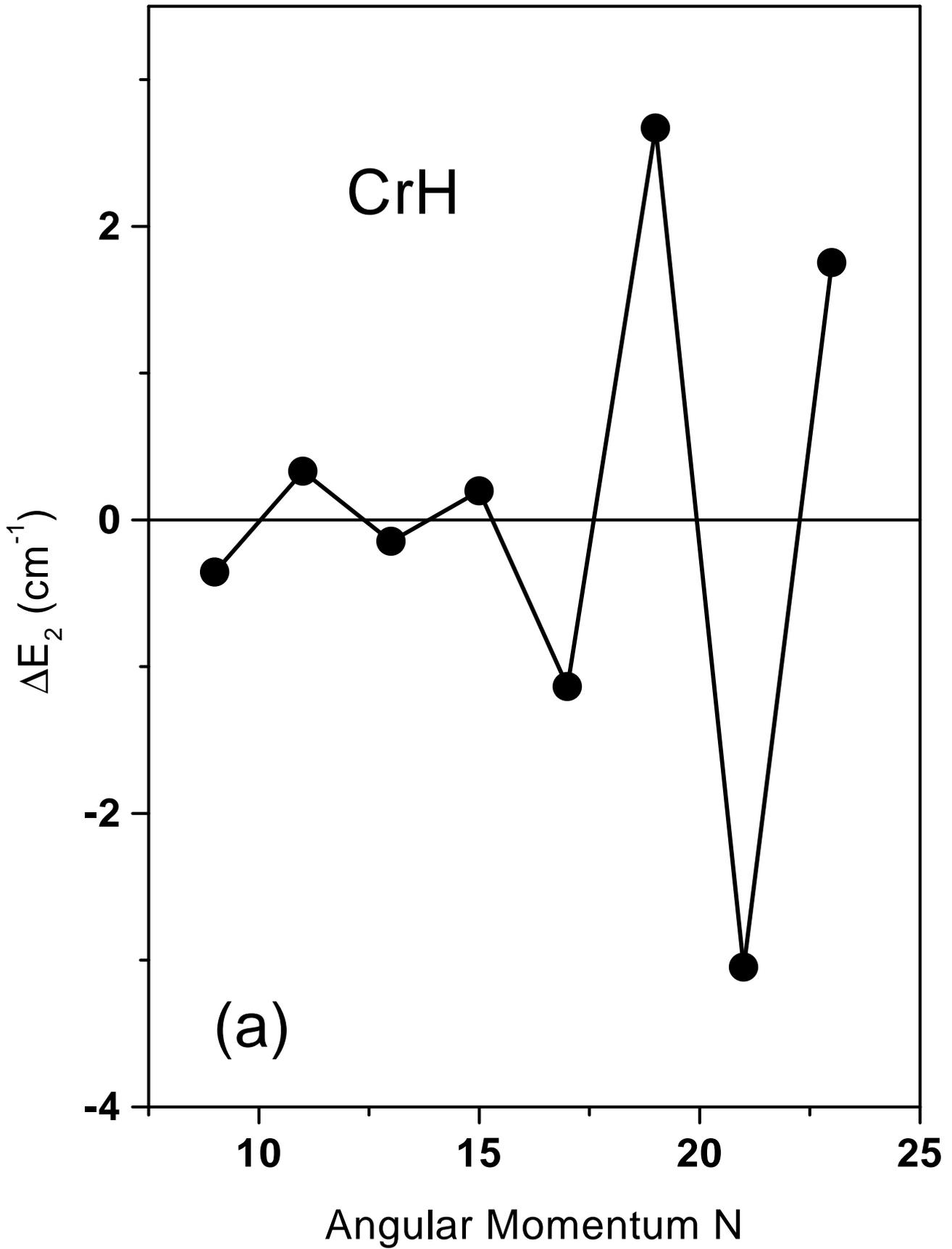
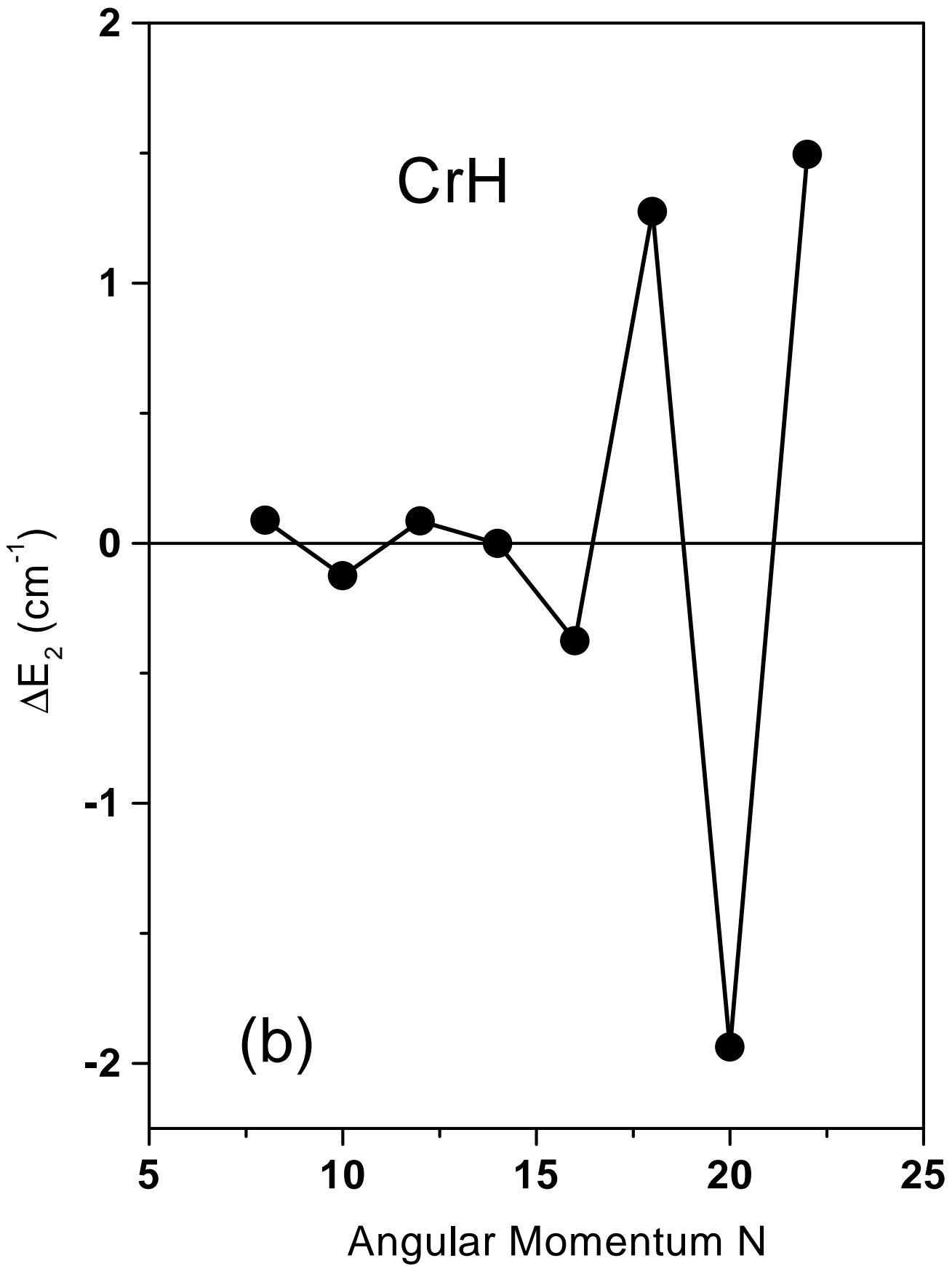
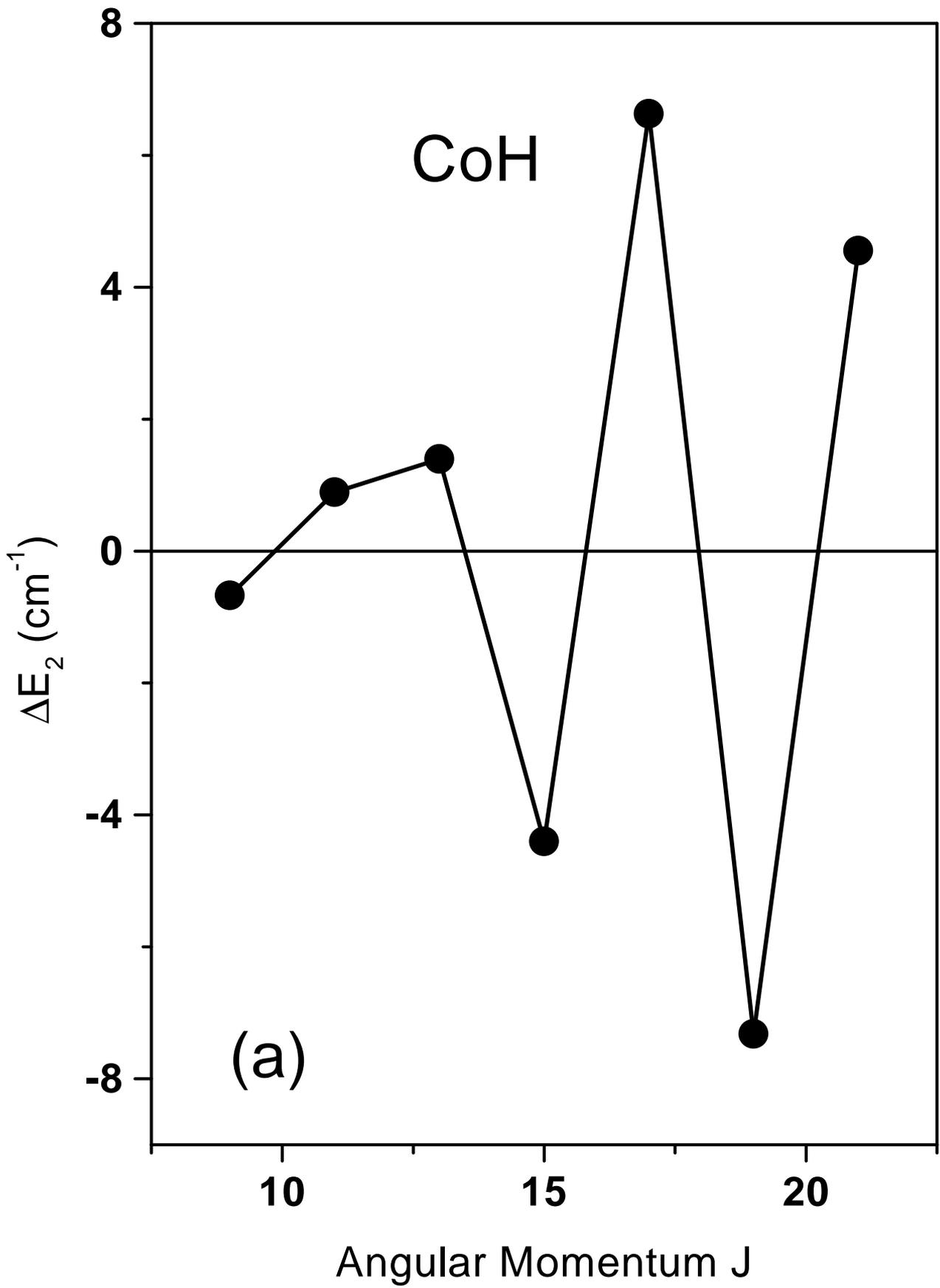
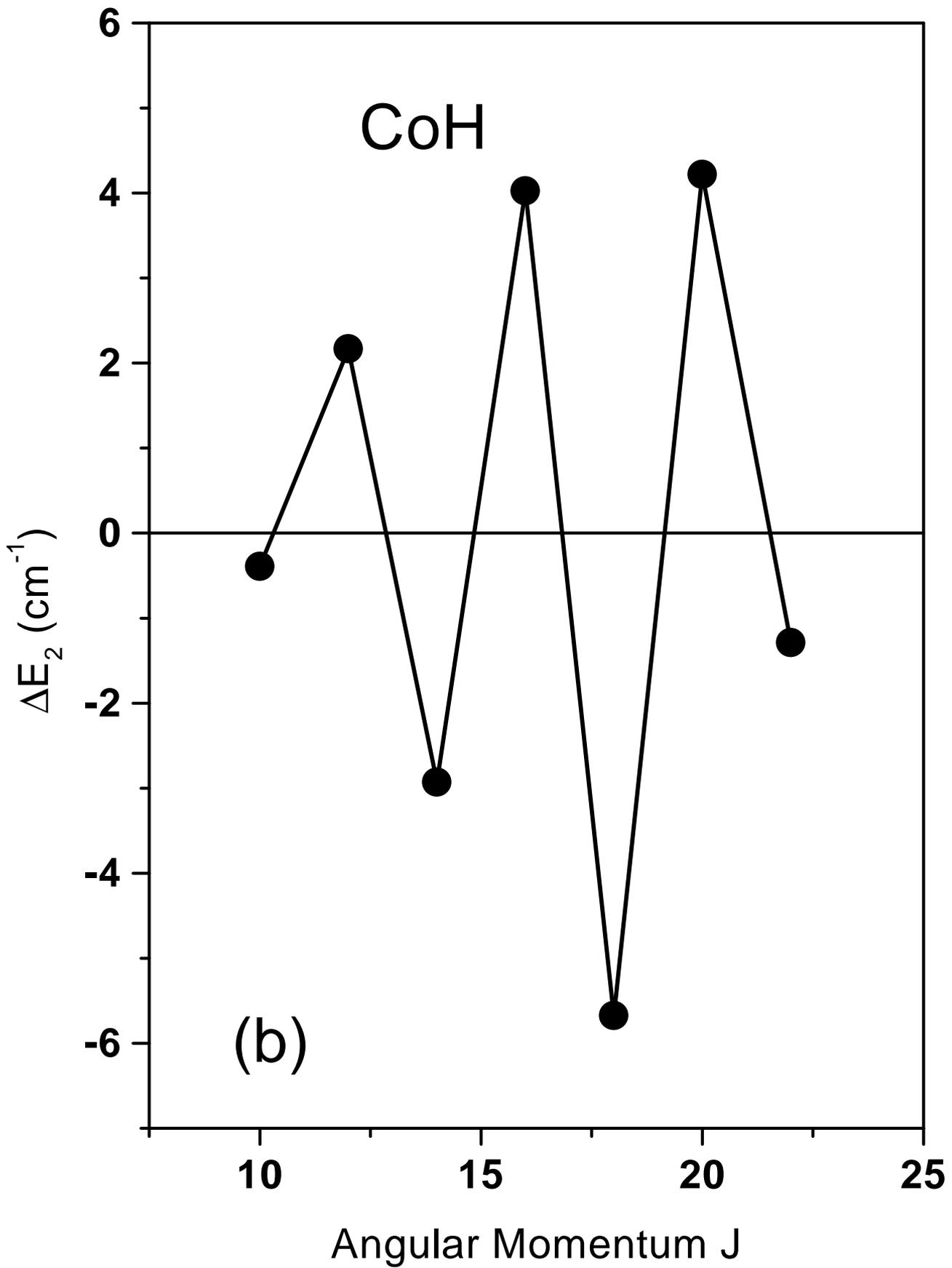
\end{document}